\documentclass[aps,a4paper,reprint,10pt,superscriptaddress,prl]{revtex4-2}
\usepackage[utf8]{inputenc}
\usepackage{amsmath}
\usepackage{amssymb}
\usepackage{braket}	
\usepackage{graphicx}
\usepackage{xcolor}
\usepackage{mathrsfs}
\usepackage[colorlinks=true, citecolor=.]{hyperref}
\usepackage{soul}
\usepackage[export]{adjustbox}
\usepackage{lipsum}
\usepackage{physics}

\newcommand{\qtr}{\operatorname{Tr}}
\newcommand{\ctr}{{\operatorfont{\mathsf{Tr}}}}

\newcommand{\btheta}{{\boldsymbol{\theta}}}
\newcommand{\hbtheta}{\skew{2.5}\hat{\boldsymbol{\theta}}}

\newtheorem{theorem}{Theorem}

\begin{document}

\title{Saturating the Quantum Cram\'er--Rao Bound in Prioritised Parameter Estimation}
\author{Simon K. Yung}
\email{sksyung@gmail.com}
\affiliation{Centre for Quantum Computation and Communication Technology, Department of Quantum Science and Technology, Research School of Physics, The Australian National University, Canberra, ACT 2601, Australia.}
\author{Aritra Das}
\affiliation{Centre for Quantum Computation and Communication Technology, Department of Quantum Science and Technology, Research School of Physics, The Australian National University, Canberra, ACT 2601, Australia.}
\author{Jun Suzuki}
\affiliation{Graduate School of Informatics and Engineering, The University of Electro-Communications, Tokyo 182-8585, Japan.}

\author{Ping Koy Lam}
\affiliation{A*STAR Quantum Innovation Centre (Q.INC), Agency for Science, Technology and Research (A*STAR), 2 Fusionopolis Way, Innovis, 138634, Singapore.}
\affiliation{Centre for Quantum Computation and Communication Technology, Department of Quantum Science and Technology, Research School of Physics, The Australian National University, Canberra, ACT 2601, Australia.}
\affiliation{Centre for Quantum Technologies, National University of Singapore, 3 Science Drive 2, Singapore 117543, Singapore.}
\author{Jie Zhao}
\email{jie.zhao@anu.edu.au}
\affiliation{Centre for Quantum Computation and Communication Technology, Department of Quantum Science and Technology, Research School of Physics, The Australian National University, Canberra, ACT 2601, Australia.}
\author{Lorc\'an O. Conlon}
\altaffiliation{Present address: Joint Quantum Institute and Joint Center for Quantum Information and Computer Science, NIST/University of Maryland, College Park, Maryland 20742, USA.}
\affiliation{A*STAR Quantum Innovation Centre (Q.INC), Agency for Science, Technology and Research (A*STAR), 2 Fusionopolis Way, Innovis, 138634, Singapore.}
\affiliation{Centre for Quantum Technologies, National University of Singapore, 3 Science Drive 2, Singapore 117543, Singapore.}
\author{Syed M. Assad}
\email{cqtsma@gmail.com}
\affiliation{A*STAR Quantum Innovation Centre (Q.INC), Agency for Science, Technology and Research (A*STAR), 2 Fusionopolis Way, Innovis, 138634, Singapore.}

\date{\today}

\begin{abstract}
	 Measurement incompatibility is a cornerstone of quantum mechanics. In the context of estimating multiple parameters of a quantum system, this manifests as a fundamental trade-off between the precisions with which different parameters can be estimated. Often, a parameter can be optimally measured, but at the cost of gaining no information about incompatible parameters. Here, we report that there are systems where one parameter's information can be maximised while not completely losing information about the other parameters. In doing so, we find attainable trade-off relations for quantum parameter estimation with a structure that is different to typical Heisenberg-type trade-offs. We demonstrate our findings by implementing an optimal entangling measurement on a Quantinuum trapped-ion quantum computer. 
\end{abstract}
\maketitle

Real physical systems rarely depend solely on a single parameter. Whether there are multiple parameters of interest or the system is influenced by unknown noise, measurement design must generally account for the system's multiparameter character. Quantum parameter estimation deals with this by providing a framework to estimate all parameters simultaneously. Due to the uncertainty principle, there is generally a trade-off between the achievable estimation errors of the parameters such that the optimal strategy must involve a compromise~\cite{liu_quantum_2019,albarelli_perspective_2020,demkowicz-dobrzanski_multi-parameter_2020,sidhu_geometric_2020}. 

It is typical to work in the local estimation framework, where the approximate values of the parameters are known and the goal is to be sensitive to small changes in the parameters. In this case, the estimation precision limit is set by the quantum Cram\'er--Rao bound (QCRB)~\cite{helstrom_minimum_1967,helstrom_minimum_1968,helstrom_quantum_1976,braunstein_statistical_1994,paris_quantum_2009}, an extension of the classical Cram\'er--Rao bound~\cite{fisher_mathematical_1922,cramer_mathematical_1946,rao_minimum_1947,rao_minimum_1947}. For estimating a single parameter, the QCRB is attainable and sets the ultimate limit on the mean squared error of the parameter's estimate~\cite{hayashi_asymptotic_2005,braunstein_statistical_1994,paris_quantum_2009}.

However, for multiparameter quantum estimation, the QCRB is generally not attainable, as the optimal measurements for estimating each parameter are typically incompatible~\cite{liu_quantum_2019,albarelli_perspective_2020,demkowicz-dobrzanski_multi-parameter_2020,sidhu_geometric_2020,matsumoto_new_2002,ragy_compatibility_2016}. In particular, a measurement that is optimal for one parameter (i.e., saturating its QCRB) may not obtain any information about the other parameters. This is the case, for example, for single qubit systems \footnote{Any projective measurement on a qubit system will produce a singular classical Fisher information matrix, because there are two dependent outcomes but more than one parameter~\cite{candeloro_dimension_2024}.}. In these cases, the parameters have been called ``informationally exclusive''~\cite{hayashi_geometrical_2005}. 

We pose the question: is it possible to be maximally sensitive to one parameter while still being sensitive to the remaining parameters? In other words, is it possible to saturate the QCRB for one parameter while learning about the other parameters? This is relevant in realistic metrology applications, for example, those afflicted by noise. There are cases where the answer is trivial, for example, when the parameters are compatible (i.e., the optimal measurements for each parameter commute). However, we are interested in situations where all parameters cannot be simultaneously optimally estimated, i.e., where there is a non-trivial estimation error trade-off. Here, and in the rest of this work, by ``maximally sensitive'' or ``optimally estimating the parameter $\theta$'' we mean that the mean squared error for the parameter $\theta$ attains its single-parameter QCRB. 

As an example, consider an ensemble of two-level atoms in Ramsey interferometry (e.g., for precision spectroscopy in atomic clocks~\cite{huelga_improvement_1997}). The atoms are prepared in the superposition $(\ket{0}+\ket{1})/\sqrt{2}$, and a phase shift, $\phi$, accumulates over time, transforming the state to $(\ket{0}+e^{i\phi}\ket{1})/\sqrt{2}$. However, the state typically also decoheres through a dephasing process that leads to a decay of the off-diagonal term of the density operator. A similar process occurs in optical interferometry, where phase estimation is affected by phase diffusion~\cite{genoni_optical_2011,genoni_optical_2012,vidrighin_joint_2014}. In either case, the phase shift is the only parameter of interest, but its precise estimation requires accurate knowledge of the decoherence strength. It is therefore natural to want to optimally estimate the phase while also learning about the decoherence.

\begin{figure}
	\includegraphics[left,width=0.95\linewidth]{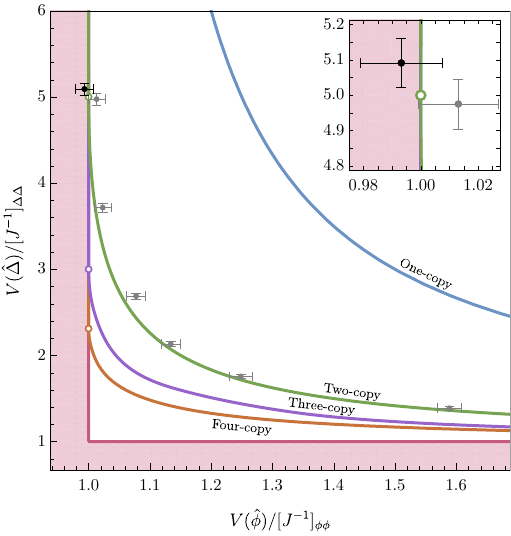}
	\caption{Trade-off curves for qubit phase--dephasing estimation, which define the boundary of the accessible region for mean squared errors. The blue, green, purple, and orange lines are the trade-off curves for one-, two-, three-, and four-copy collective measurements, respectively. The mean squared errors are scaled by respective quantum Fisher informations, so that a value of 1 represents optimal estimation. The boundary of the shaded region is the quantum Cram\'er--Rao bound and is attainable in the infinite-copy limit for this system. The unfilled circles on the curves denote the minimum mean squared errors for phase-prioritised estimation. The black (grey) points are the mean squared errors from the trapped-ion experiment (emulated experiment), with error bars denoting one standard deviation obtained via bootstrapping. The true values of the parameters are $\phi=0$ and $\Delta=1/2$, but these values do not affect the features of the trade-off. Inset: magnified plot centred around the optimal two-copy phase-prioritied mean squared errors.}
	\label{fig:trade-off-expresults}
\end{figure}

In the case of phase--dephasing estimation with measurements performed on individual qubits, any measurement that is maximally sensitive to the phase is not sensitive to the dephasing strength. This is exemplified by the blue line in Fig.~\ref{fig:trade-off-expresults}, where we plot the fundamental trade-off between the mean squared error of the phase estimate ($\hat{\phi}$) and the dephasing strength estimate ($\hat{\Delta}$), i.e., error-pairs below the line are impossible when measuring individual qubits. However, we could alternatively perform collective measurements on pairs of qubits, which leads to a different accessible region (see green line in Fig.~\ref{fig:trade-off-expresults}). In particular, with two-copy measurements, the phase can be optimally estimated while learning about the dephasing. However, the converse is not true: a measurement that optimally estimates the dephasing is not sensitive to the phase. 

In the remainder of this Letter, we reveal the solution to our question: there is a straightforward method to check whether or not a parameter can be optimally estimated while being sensitive to the other parameters. When there are only two parameters, we present a technique to directly identify the optimal prioritised measurement. We further describe the possible generalisations to more than two parameters. We begin with a brief overview of the quantum parameter estimation framework.

\textit{Quantum parameter estimation}---Consider a density operator $\rho(\btheta)$ parameterised by real parameters $\btheta = (\theta_1,\dots,\theta_m)^\top$. We may obtain an estimate of the parameters, $\hbtheta$, by performing measurements described by a positive operator-valued measure (POVM) $\Pi = \{\Pi_i \ | \ \Pi_i\succeq 0, \sum_i \Pi_i = \openone\}$. We work in the local estimation framework, where the true values of the parameters are approximately known, and enforce that the estimator is locally unbiased: $\mathbb{E}[\hbtheta] = \btheta$ and $\partial\mathbb{E}[\hat{\theta}_i]/\partial \theta_j = \delta_{ij}$ at the true parameter values, where the expectation value is taken over the probability distribution $p(k|\btheta) =\qtr[\Pi_k\rho]$. The performance of the measurement-estimation process is quantified by the mean squared error matrix $[V(\hbtheta,\Pi)]_{jk} = \mathbb{E}[(\hat{\theta}_j-\theta_j)(\hat{\theta}_k-\theta_k)]$. 

For a particular measurement, the mean squared error is bounded by the classical Fisher information, $F$, of the probability distribution $p(k|\btheta)$: $V(\hbtheta,\Pi) \succeq F^{-1}/N$; this bound is attainable in the asymptotic limit of number of samples ($N$), e.g., by maximum likelihood estimation~\cite{fisher_mathematical_1922,cramer_mathematical_1946,rao_minimum_1947,rao_minimum_1947}. The factor of $1/N$ quantifies the scaling with resources used, which we hereafter suppress, equivalent to considering the scaled mean squared error $NV(\hbtheta,\Pi)$. The classical Fisher information is in turn bounded by the quantum Fisher information, $J=J(\rho,\theta)$, which only depends on the state's density operator and its derivatives \cite{helstrom_minimum_1967,helstrom_minimum_1968,helstrom_quantum_1976,braunstein_statistical_1994,paris_quantum_2009}. In the local framework, we evaluate the Fisher informations at the true parameter values.

In particular, the scalar lower bound $\ctr[W V(\hbtheta,\Pi)] \geq \ctr[WJ^{-1}]$, for a positive definite matrix $W$, is called the quantum Cram\'er--Rao bound. A consequence is that the mean squared error for each parameter is limited by the corresponding diagonal element of $J^{-1}$. For single-parameter estimation, the quantum Cram\'er--Rao bound is attainable by a projective measurement in the eigenbasis of the symmetric logarithmic derivative (SLD) operator $L_{\theta_j}$, defined implicitly by $\partial\rho/\partial\theta_j = (L_{\theta_j} \rho+\rho L_{\theta_j})/2$. However, for multiparameter estimation, the SLD operators for different parameters typically do not commute, and, as a result, the quantum Cram\'er--Rao bound is generally not attainable. Instead, tighter lower bounds have been developed that take the measurement incompatibility into account~\cite{yuen_multiple-parameter_1973,gill_state_2000,holevo_statistical_1973,holevo_probabilistic_2011,nagaoka_new_2005,nagaoka_generalization_2005,conlon_efficient_2021,lu_incorporating_2021}. In this work, we use the Nagaoka Cram\'er--Rao bound, which applies for two-parameter estimation~\cite{nagaoka_new_2005,nagaoka_generalization_2005}. Sometimes, this bound coincides with the trade-off relation developed in Ref.~\cite{lu_incorporating_2021}, which can simplify calculations of the trade-off curves~\cite{yung_comparison_2024}. 

\textit{Phase-estimation with unknown dephasing}---Here, we consider the aforementioned phase--dephasing estimation problem in detail. We suppose that the pure qubit state $(\ket{0}+\ket{1})(\bra{0}+\bra{1})/2$ undergoes a phase shift $\rho \rightarrow e^{-i\phi\sigma_z/2}\rho e^{i\phi\sigma_z/2}$ and dephasing $\rho \rightarrow (1-\Delta/2)\rho  + (\Delta/2) \sigma_z\rho\sigma_z$, where the phase shift $\phi$ and the dephasing strength $\Delta$ are the parameters to be estimated, and $\sigma_z=\ketbra{0}-\ketbra{1}$. This transformation can be considered as the result of a channel that applies a random phase shift, distributed normally with mean $\phi$ and variance $2\ln(1/(1-\Delta))$. 

For this problem, the SLD operator for $\phi$ is $L_\phi = i(1-\Delta)(e^{-i\phi}\ketbra{1}{0}-e^{i\phi}\ketbra{0}{1})$. 
The dependence of $L_\phi$ on $\Delta$ means that the optimal procedure to estimate $\phi$ depends on the true value of $\Delta$, specifically, the post-processing of measurement outcomes. However, the corresponding projective measurement in the $L_\phi$-eigenbasis does not gain any information about $\Delta$ (only one parameter can be estimated from a two-outcome measurement). As such, the $\phi$-optimal measurement is not amenable to an adaptive implementation as it is not sensitive to changes in $\Delta$. The estimation error trade-off is exemplified in Fig.~\ref{fig:trade-off-expresults}, where minimising the mean squared error for $\phi$ results in a diverging mean squared error for $\Delta$, and vice versa (blue line). See Supplemental Material for details on the trade-off calculation \footnote{See Supplemental Material for additional details on the phase--dephasing estimation problem and the Fock state displacement sensing problem, details on the Quantinuum experiment, and the proof of Theorem 1.}.

As previously mentioned, the estimation error trade-off can be partly mitigated by implementing a collective measurement on multiple copies of the quantum state. The trade-off curve for two-copy measurements is shown in Fig.~\ref{fig:trade-off-expresults} (green line), calculated from the Nagaoka Cram\'er--Rao bound for the state $\rho^{\otimes 2} \equiv \rho\otimes \rho$. In particular, as the mean squared error for $\phi$ is reduced to its minimum ($[J^{-1}]_{\phi\phi}$, as determined by the quantum Cram\'er--Rao bound with $W=(1,0)^\top (1,0)$), the mean squared error for $\Delta$ can take finite values (although the reverse is not true). In this case, we can attribute this to the extra freedom in choosing measurements performed on two copies of the quantum state, i.e., using entanglement. The two-copy SLD operators are $L_\phi^{(2)} = L_\phi \otimes \openone + \openone\otimes L_\phi$, which have degenerate eigenvalues. There is thus some freedom in selecting an eigenbasis of $L_\phi^{(2)}$ to measure in, and some choices may provide information about the other parameter, $\Delta$. This is exactly what we find for $L_\phi^{(2)}$. 

Explicitly, let $\ket{l_+}$ and $\ket{l_-}$ be the eigenvectors of $L_\phi$ (with eigenvalues $\pm (1-\Delta)$). Then, the eigenspaces of $L_\phi^{(2)}$ partition the two-qubit Hilbert space based on the projectors $P_0 = \ket{l_-}\ket{l_+}\bra{l_-}\bra{l_+}+\ket{l_+}\ket{l_-}\bra{l_+}\bra{l_-}$ and $P_\pm = \ket{l_\pm}\ket{l_\pm}\bra{l_\pm}\bra{l_\pm}$, with eigenvalues $0$ and $\pm 2(1-\Delta)$, respectively. The projective measurement $\{P_0,P_-,P_+\}$ saturates the quantum Cram\'er--Rao bound for estimating only $\phi$, in the sense that the scaled mean squared error for $\hat\phi$ is equal to the inverse of the quantum Fisher information about $\phi$. In fact, $P_0$ does not give any information about $\phi$ as the corresponding eigenvalue of $L_\phi^{(2)}$ is zero. 

Similarly, the projectors $P_\pm$ provide no information about $\Delta$, which can be seen as the projected states $P_\pm\rho^{\otimes 2} P_\pm$ do not depend on $\Delta$. However, $P_0\rho^{\otimes 2} P_0$ does depend on $\Delta$. This means that a judicious choice of fine-grained measurement within the subspace $P_0$ allows us to learn about $\Delta$ while not affecting the optimality of the $\phi$-estimate. An optimal fine-grained measurement (i.e., one that maximises the $\Delta$ information) can be determined by the SLD operator for $\Delta$ of the projected state (see Supplemental Material for details~\cite{Note2}). We call this the two-copy $\phi$-prioritised optimal measurement. Its mean squared error coincides with the unfilled green point in Fig.~\ref{fig:trade-off-expresults}, where the two-copy trade-off curve meets the vertical line of the quantum Cram\'er--Rao bound. On the other hand, the two-copy state projected onto the eigenspaces of $L_\Delta^{(2)}$ yield zero quantum Fisher information about $\phi$. 
 
We experimentally demonstrate the features of the two-copy phase--dephasing estimation trade-off using a trapped-ion quantum processor (Quantinuum H1-1~\footnote{Quantinuum H1-1, \url{https://quantinuum.com/}, accessed April 10--11 2025.}). We implemented the two-copy $\phi$-prioritised optimal measurement on the actual hardware and emulated other optimal measurements along the curve, whose POVMs were found using numerical optimisation. The results are presented in Fig.~\ref{fig:trade-off-expresults}, showing good agreement with the theoretical limit. This demonstrates two-copy entangling measurements attaining the theoretical precision limit despite the noise limitations of current state-of-the-art hardware. In addition, the closeness of the experimental and emulated results for the $\phi$-prioritised measurement suggests that the results of experimentally implemented measurements along the curve would be similar to the emulated results. Details on the experiment are provided in the Supplemental Material \cite{Note2}. 

\textit{Prioritised parameter estimation}---We call the feature seen above, where the mean squared error of one parameter can saturate the quantum Cram\'er--Rao bound with finite mean squared error for the other parameters, \textit{prioritised parameter estimation}. This is practically relevant to systems where only one parameter is of actual interest, but its precise and accurate measurement also depends on other parameters. For now, we assume that the state depends on only two parameters: one priority parameter, $\theta_p$, and one other parameter, $\theta_o$. Generalisations to more parameters will be discussed later.   

Ultimately, prioritised parameter estimation is afforded by a partial compatibility of optimal measurements. Specifically, we consider a measurement that saturates the quantum Cram\'er--Rao bound for $\theta_p$, described by the projectors $\{P_j\}$ onto eigenspaces of the corresponding observable, $O_p$. Such a measurement may be sensitive to $\theta_o$, or can be refined to be sensitive to $\theta_o$ if the state projected onto any of the eigenspaces depends on $\theta_o$ (e.g., if the rank of a projector is greater than 1). If this is the case, the outcome probabilities of a projective measurement in the $O_p$ eigenbasis depend on $\theta_o$, allowing an estimate to be formed. We formulate this as the following theorem that can be applied straightforwardly to any full-rank two-parameter system. 
\begin{theorem}
	Let $\rho(\theta_p,\theta_o)$ be a regular full-rank two-parameter quantum statistical model with linearly independent parameters. Let $O_p$ be the optimal observable for estimating $\theta_p$. Then, $\theta_o$ can be estimated whilst optimally estimating $\theta_p$ if and only if the projectors $\{P_j\}$ of $O_p$ can be chosen such that $\partial_{\theta_o}\qtr[\rho_\theta P_j]\neq 0$ for some $j$. Furthermore, this is equivalent to the classical Fisher information of the projective measurement $\{P_j\}$ being positive-definite. 
\end{theorem}

The optimal observable $O_p$ is the SLD operator $L_p$ when $J(\rho,\btheta)$ is diagonal, and otherwise a linear combination of the SLD operators~\cite{suzuki_quantum_2020}. We provide the proof of the theorem in the Supplemental Material \cite{Note2}. When the model is rank-deficient, as the SLD operator is not unique, a search over all SLD operators is generally required. We remark that this two-stage measurement design is distinct from post-selected quantum metrology~\cite{arvidsson-shukur_quantum_2020,das_saturating_2023}, where a measurement is performed to select states and conditionally increase the quantum Fisher information, because we also utilise the coarse measurement to estimate the prioritised parameter.  

We note that the phase--dephasing example is interesting in the regard that collective measurements are required for phase-prioritised estimation. Collective measurements, however, are neither necessary nor sufficient in general. To prove this for necessity, note that the two-copy qubit system can be considered as a single four-dimensional system without collective measurements. For sufficiency, note that dephasing-prioritised estimation is not possible even with collective measurements on any finite number of copies of the state. The latter argument also indicates that the degeneracy of an SLD eigenspace is not a sufficient condition for prioritised estimation. Such degeneracy is also not necessary, as prioritised estimation would be possible in a system in which the SLD operators share an eigenvector (with non-zero eigenvalues). Below, we present an example of a system where prioritised parameter estimation is possible for either parameter.         

\textit{Further examples}---We consider sensing the conjugate parameters of a displacement channel $D(\theta_x,\theta_y) =\exp{(i\theta_x\hat{y}+ i\theta_y\hat{x})}$. The channel displaces a single-mode optical field with annihilation and creation operators $\hat{a},\hat{a}^\dagger$ by $\theta_x$ in the amplitude quadrature $\hat{x}=(\hat{a}^\dagger+\hat{a})/\sqrt{2}$ and $\theta_y$ in the phase quadrature $\hat{y} = i(\hat{a}^\dagger-\hat{a})/\sqrt{2}$. For simplicity, we assume the true values $\theta_x=\theta_y=0$ and probe the channel with Fock states $\ket{n}$ satisfying $\hat{a}^\dagger\hat{a}\ket{n} = n\ket{n}$ for $n\geq 0$. For $n=0$, the estimation error trade-off was studied in Ref.~\cite{lu_incorporating_2021}.

\begin{figure}
	\centering
	\includegraphics[width=0.975\linewidth]{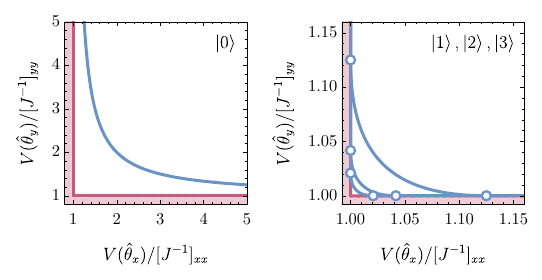}	
	\caption{Estimation error trade-off curves for displacement sensing with Fock states $\ket{n}$ (curves for $\ket{1}$, $\ket{2}$, $\ket{3}$ in order towards bottom-left corner). The mean squared errors are normalised by the respective quantum Fisher informations (which are different for each $n$) so that the axes represent the closeness to optimal for each parameter and the trade-offs can be compared. The unfilled circles denote the minimum achievable prioritised mean squared errors. The boundary of the shaded region is the quantum Cram\'er--Rao bound.}
	\label{fig:disptradeoff}
\end{figure}

 The trade-off curves for different probe states $\ket{n}$ are presented in Fig.~\ref{fig:disptradeoff}. There is a clear qualitative difference between the trade-offs for $n=0$ and $n>0$. The informational exclusiveness for $n=0$ can be understood by considering that the optimal way to estimate $\theta_x$ is to measure $\hat{x}$, which reveals nothing about $\hat{y}$ (and $\theta_y$).

In Fock state displacement sensing with $n\geq 1$, prioritised parameter estimation is possible for either $\theta_x$ or $\theta_y$. In this case, the SLD operators do not have degenerate eigenspaces, but there is still freedom to choose the optimal projective measurement because the SLD operators are not unique for rank-deficient states. A measurement in the eigenbasis of a particular choice of SLD operator can saturate one of the open circles in Fig.~\ref{fig:disptradeoff} (see Supplemental Material for details \cite{Note2}).

We also note that the possibility of prioritised parameter estimation can be observed in studies of other physical problems, for example, displacement sensing with two-mode Gaussian probes~\cite{assad_accessible_2020,bradshaw_tight_2017,bradshaw_ultimate_2018}, linear measurements for gravitational wave detection~\cite{gardner_achieving_2024,li_general_2024}, the estimation of spatial displacements and tilts of light using Hermite--Gaussian beams~\cite{xia_toward_2023}, and simultaneous phase and loss estimation~\cite{crowley_tradeoff_2014}. 

\textit{Discussion}---So far, we have only considered systems depending on two parameters. With more than two parameters, there may be several generalisations of prioritised parameter estimation, depending on the situation and the structure of the estimation error trade-off. We briefly discuss potential generalisations, but leave a more complete analysis to future work. 

The simplest generalisation is a scenario with a single priority parameter and several other parameters, for example, if there were multiple sources of noise. In this case, the method of projecting onto SLD eigenbases, with the addition of applying multiparameter estimation techniques to the projected states, e.g., applying the Nagaoka--Hayashi Cram\'er--Rao bound, is suitable. We also envision generalisations with multiple priority parameters, or different degrees of priority for the different parameters. For example, with three parameters, we could aim to optimally estimate $\theta_1$ while being sensitive to $\theta_2$ and $\theta_3$. Then, if the $\theta_1$-optimal measurement is not unique, we could tune the measurement to increase the sensitivity to $\theta_2$ without losing all information about $\theta_3$. 

We note that prioritised parameter estimation can be considered a variant of nuisance parameter estimation~\cite{suzuki_nuisance_2020,suzuki_quantum_2020}, in which a subset of parameters are not of interest but must nevertheless be accounted for (e.g., in parameter correlations or the optimal measurements). However, while information about the nuisance parameters is not sought, their approximate values must typically be known. This means that, practically, a measurement must first be performed to determine all parameters to within some precision. A more efficient measurement can then be made to better estimate the parameters of interest, but its performance will remain limited by the precision of the first measurement. Where prioritised parameter estimation is possible, this drawback is resolved, and an adaptive scheme can be implemented to optimally estimate the parameters of interest.    

Finally, we discuss a possible application of the prioritised estimation approach to finding near-optimal collective measurements for ``standard'' parameter estimation (i.e., for minimising $\ctr[V(\hat\btheta,\Pi)]$). Due to the large Hilbert space of a multiple-copy probe state, it is generally difficult to find optimal measurements that minimise the mean squared error sum, as brute-force numerical methods must typically be used when there is measurement incompatibility. This poses a challenge for the practical implementation of collective measurements to obtain a quantum advantage. This is exemplified in the relatively small number of non-trivial optimal collective measurements that have been determined and implemented for physical estimation problems~\cite{hou_deterministic_2018,conlon_approaching_2023,conlon_discriminating_2023,zhou_experimental_2025}, despite the advantage of collective measurements being known for at least three decades~\cite{massar_optimal_1995}. 

\begin{figure}
	\centering
	\includegraphics{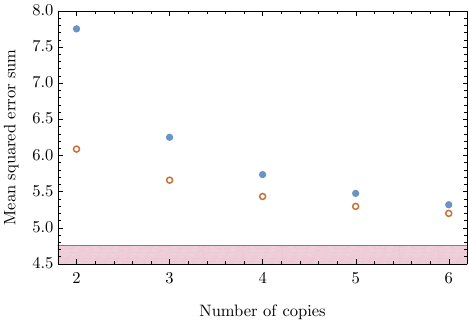}
	\caption{Sum of mean squared errors $V(\hat{\phi})+V(\hat{\Delta})$ of $\phi$-prioritised measurements (blue, filled circles) with different numbers of copies of the probe state. The orange unfilled circles are the Nagaoka--Cram\'er--Rao bound, a lower bound for all measurements, and the boundary of the pink is the quantum Cram\'er--Rao bound which here is attainable in the asymptotic limit of the number of copies. The mean squared errors are scaled by the number of probe state copies used. The true values of the parameters are $\phi=0$ and $\Delta=1/2$.}
	\label{fig:Ncopies}
\end{figure}

To elucidate the utility of the prioritised estimation approach for near-optimal collective measurements, we revisit the phase--dephasing estimation problem. In Fig.~\ref{fig:trade-off-expresults}, it is evident that the phase-prioritised measurement achieves a mean squared error that approaches the infinite-copy limit as the number of copies increases. The mean squared error sum of such measurements also approaches the Nagaoka Cram\'er--Rao bound with the same number of copies, as shown in Fig.~\ref{fig:Ncopies}. Here, as the number of copies increases, the prioritised measurement achieves near-optimal precision. In addition, the measurement does not become appreciably more complex to determine---the method of projecting onto the SLD operator eigenspaces works for any number of copies and does not require any optimisation. 

\textit{Conclusions}---We have studied the structure of the estimation error trade-off in estimation problems without complete incompatibility between parameters. This has highlighted that in quantum multiparameter estimation, there is an interesting middle-ground between compatible parameters (i.e., commuting SLD operators) and informationally exclusive parameters. In particular, there is a regime where the quantum Cram\'er--Rao bound for one parameter can be saturated while learning about another parameter, exhibiting a trade-off structure that is starkly different to typical Heisenberg-type trade-offs. 

In studying prioritised parameter estimation, we have presented general analytic conditions for completely determining optimal single-parameter prioritised measurements, without the need for complex optimisation. We have implemented an optimal entangling measurement for estimating a phase shift in the presence of unknown dephasing and demonstrated that similar collective measurements on many copies of the probe state can be both near-optimal and simple to determine.  

\textit{Acknowledgements}---This research was supported by the Australian Research Council Centre of Excellence CE170100012. This research was also supported by A*STAR C230917010, Emerging Technology and A*STAR C230917004, Quantum Sensing. S.K.Y. is supported by the Australian Government Research Training Program Scholarship. J.S. is supported by JSPS KAKENHI Grant Numbers JP24K14816 and ERATO ``Super Quantum Entanglement" (Grant No. JPMJER2402) from JST.

\bibliography{bib}

\begin{thebibliography}{49}%
\makeatletter
\providecommand \@ifxundefined [1]{%
 \@ifx{#1\undefined}
}%
\providecommand \@ifnum [1]{%
 \ifnum #1\expandafter \@firstoftwo
 \else \expandafter \@secondoftwo
 \fi
}%
\providecommand \@ifx [1]{%
 \ifx #1\expandafter \@firstoftwo
 \else \expandafter \@secondoftwo
 \fi
}%
\providecommand \natexlab [1]{#1}%
\providecommand \enquote  [1]{``#1''}%
\providecommand \bibnamefont  [1]{#1}%
\providecommand \bibfnamefont [1]{#1}%
\providecommand \citenamefont [1]{#1}%
\providecommand \href@noop [0]{\@secondoftwo}%
\providecommand \href [0]{\begingroup \@sanitize@url \@href}%
\providecommand \@href[1]{\@@startlink{#1}\@@href}%
\providecommand \@@href[1]{\endgroup#1\@@endlink}%
\providecommand \@sanitize@url [0]{\catcode `\\12\catcode `\$12\catcode `\&12\catcode `\#12\catcode `\^12\catcode `\_12\catcode `\%12\relax}%
\providecommand \@@startlink[1]{}%
\providecommand \@@endlink[0]{}%
\providecommand \url  [0]{\begingroup\@sanitize@url \@url }%
\providecommand \@url [1]{\endgroup\@href {#1}{\urlprefix }}%
\providecommand \urlprefix  [0]{URL }%
\providecommand \Eprint [0]{\href }%
\providecommand \doibase [0]{https://doi.org/}%
\providecommand \selectlanguage [0]{\@gobble}%
\providecommand \bibinfo  [0]{\@secondoftwo}%
\providecommand \bibfield  [0]{\@secondoftwo}%
\providecommand \translation [1]{[#1]}%
\providecommand \BibitemOpen [0]{}%
\providecommand \bibitemStop [0]{}%
\providecommand \bibitemNoStop [0]{.\EOS\space}%
\providecommand \EOS [0]{\spacefactor3000\relax}%
\providecommand \BibitemShut  [1]{\csname bibitem#1\endcsname}%
\let\auto@bib@innerbib\@empty
\bibitem [{\citenamefont {Liu}\ \emph {et~al.}(2019)\citenamefont {Liu}, \citenamefont {Yuan}, \citenamefont {Lu},\ and\ \citenamefont {Wang}}]{liu_quantum_2019}%
  \BibitemOpen
  \bibfield  {author} {\bibinfo {author} {\bibfnamefont {J.}~\bibnamefont {Liu}}, \bibinfo {author} {\bibfnamefont {H.}~\bibnamefont {Yuan}}, \bibinfo {author} {\bibfnamefont {X.-M.}\ \bibnamefont {Lu}},\ and\ \bibinfo {author} {\bibfnamefont {X.}~\bibnamefont {Wang}},\ }\bibfield  {title} {\bibinfo {title} {Quantum {Fisher} information matrix and multiparameter estimation},\ }\href {https://doi.org/10.1088/1751-8121/ab5d4d} {\bibfield  {journal} {\bibinfo  {journal} {Journal of Physics A: Mathematical and Theoretical}\ }\textbf {\bibinfo {volume} {53}},\ \bibinfo {pages} {023001} (\bibinfo {year} {2019})}\BibitemShut {NoStop}%
\bibitem [{\citenamefont {Albarelli}\ \emph {et~al.}(2020)\citenamefont {Albarelli}, \citenamefont {Barbieri}, \citenamefont {Genoni},\ and\ \citenamefont {Gianani}}]{albarelli_perspective_2020}%
  \BibitemOpen
  \bibfield  {author} {\bibinfo {author} {\bibfnamefont {F.}~\bibnamefont {Albarelli}}, \bibinfo {author} {\bibfnamefont {M.}~\bibnamefont {Barbieri}}, \bibinfo {author} {\bibfnamefont {M.}~\bibnamefont {Genoni}},\ and\ \bibinfo {author} {\bibfnamefont {I.}~\bibnamefont {Gianani}},\ }\bibfield  {title} {\bibinfo {title} {A perspective on multiparameter quantum metrology: {From} theoretical tools to applications in quantum imaging},\ }\href {https://doi.org/10.1016/j.physleta.2020.126311} {\bibfield  {journal} {\bibinfo  {journal} {Physics Letters A}\ }\textbf {\bibinfo {volume} {384}},\ \bibinfo {pages} {126311} (\bibinfo {year} {2020})}\BibitemShut {NoStop}%
\bibitem [{\citenamefont {Demkowicz-Dobrza{\'n}ski}\ \emph {et~al.}(2020)\citenamefont {Demkowicz-Dobrza{\'n}ski}, \citenamefont {G{\'o}recki},\ and\ \citenamefont {Gu{\c t}{\u a}}}]{demkowicz-dobrzanski_multi-parameter_2020}%
  \BibitemOpen
  \bibfield  {author} {\bibinfo {author} {\bibfnamefont {R.}~\bibnamefont {Demkowicz-Dobrza{\'n}ski}}, \bibinfo {author} {\bibfnamefont {W.}~\bibnamefont {G{\'o}recki}},\ and\ \bibinfo {author} {\bibfnamefont {M.}~\bibnamefont {Gu{\c t}{\u a}}},\ }\bibfield  {title} {\bibinfo {title} {Multi-parameter estimation beyond quantum {Fisher} information},\ }\href {https://doi.org/10.1088/1751-8121/ab8ef3} {\bibfield  {journal} {\bibinfo  {journal} {Journal of Physics A: Mathematical and Theoretical}\ }\textbf {\bibinfo {volume} {53}},\ \bibinfo {pages} {363001} (\bibinfo {year} {2020})}\BibitemShut {NoStop}%
\bibitem [{\citenamefont {Sidhu}\ and\ \citenamefont {Kok}(2020)}]{sidhu_geometric_2020}%
  \BibitemOpen
  \bibfield  {author} {\bibinfo {author} {\bibfnamefont {J.~S.}\ \bibnamefont {Sidhu}}\ and\ \bibinfo {author} {\bibfnamefont {P.}~\bibnamefont {Kok}},\ }\bibfield  {title} {\bibinfo {title} {Geometric perspective on quantum parameter estimation},\ }\href {https://doi.org/10.1116/1.5119961} {\bibfield  {journal} {\bibinfo  {journal} {AVS Quantum Science}\ }\textbf {\bibinfo {volume} {2}},\ \bibinfo {pages} {014701} (\bibinfo {year} {2020})}\BibitemShut {NoStop}%
\bibitem [{\citenamefont {Helstrom}(1967)}]{helstrom_minimum_1967}%
  \BibitemOpen
  \bibfield  {author} {\bibinfo {author} {\bibfnamefont {C.~W.}\ \bibnamefont {Helstrom}},\ }\bibfield  {title} {\bibinfo {title} {Minimum mean-squared error of estimates in quantum statistics},\ }\href {https://doi.org/10.1016/0375-9601(67)90366-0} {\bibfield  {journal} {\bibinfo  {journal} {Physics Letters A}\ }\textbf {\bibinfo {volume} {25}},\ \bibinfo {pages} {101} (\bibinfo {year} {1967})}\BibitemShut {NoStop}%
\bibitem [{\citenamefont {Helstrom}(1968)}]{helstrom_minimum_1968}%
  \BibitemOpen
  \bibfield  {author} {\bibinfo {author} {\bibfnamefont {C.}~\bibnamefont {Helstrom}},\ }\bibfield  {title} {\bibinfo {title} {The minimum variance of estimates in quantum signal detection},\ }\href {https://doi.org/10.1109/TIT.1968.1054108} {\bibfield  {journal} {\bibinfo  {journal} {IEEE Transactions on Information Theory}\ }\textbf {\bibinfo {volume} {14}},\ \bibinfo {pages} {234} (\bibinfo {year} {1968})}\BibitemShut {NoStop}%
\bibitem [{\citenamefont {Helstrom}(1976)}]{helstrom_quantum_1976}%
  \BibitemOpen
  \bibfield  {author} {\bibinfo {author} {\bibfnamefont {C.~W.}\ \bibnamefont {Helstrom}},\ }\href@noop {} {\emph {\bibinfo {title} {Quantum detection and estimation theory}}},\ \bibinfo {series} {Mathematics in science and engineering}\ No.\ \bibinfo {number} {v. 123}\ (\bibinfo  {publisher} {Academic Press},\ \bibinfo {address} {New York},\ \bibinfo {year} {1976})\BibitemShut {NoStop}%
\bibitem [{\citenamefont {Braunstein}\ and\ \citenamefont {Caves}(1994)}]{braunstein_statistical_1994}%
  \BibitemOpen
  \bibfield  {author} {\bibinfo {author} {\bibfnamefont {S.~L.}\ \bibnamefont {Braunstein}}\ and\ \bibinfo {author} {\bibfnamefont {C.~M.}\ \bibnamefont {Caves}},\ }\bibfield  {title} {\bibinfo {title} {Statistical distance and the geometry of quantum states},\ }\href {https://doi.org/10.1103/PhysRevLett.72.3439} {\bibfield  {journal} {\bibinfo  {journal} {Physical Review Letters}\ }\textbf {\bibinfo {volume} {72}},\ \bibinfo {pages} {3439} (\bibinfo {year} {1994})}\BibitemShut {NoStop}%
\bibitem [{\citenamefont {Paris}(2009)}]{paris_quantum_2009}%
  \BibitemOpen
  \bibfield  {author} {\bibinfo {author} {\bibfnamefont {M.~G.~A.}\ \bibnamefont {Paris}},\ }\bibfield  {title} {\bibinfo {title} {Quantum {Estimation} for {Quantum} {Technology}},\ }\href {https://doi.org/10.1142/S0219749909004839} {\bibfield  {journal} {\bibinfo  {journal} {International Journal of Quantum Information}\ }\textbf {\bibinfo {volume} {07}},\ \bibinfo {pages} {125} (\bibinfo {year} {2009})}\BibitemShut {NoStop}%
\bibitem [{\citenamefont {Fisher}(1922)}]{fisher_mathematical_1922}%
  \BibitemOpen
  \bibfield  {author} {\bibinfo {author} {\bibfnamefont {R.~A.}\ \bibnamefont {Fisher}},\ }\bibfield  {title} {\bibinfo {title} {On the mathematical foundations of theoretical statistics},\ }\href {https://doi.org/10.1098/rsta.1922.0009} {\bibfield  {journal} {\bibinfo  {journal} {Philosophical Transactions of the Royal Society of London. Series A, Containing Papers of a Mathematical or Physical Character}\ }\textbf {\bibinfo {volume} {222}},\ \bibinfo {pages} {309} (\bibinfo {year} {1922})}\BibitemShut {NoStop}%
\bibitem [{\citenamefont {Cram{\'e}r}(1946)}]{cramer_mathematical_1946}%
  \BibitemOpen
  \bibfield  {author} {\bibinfo {author} {\bibfnamefont {H.}~\bibnamefont {Cram{\'e}r}},\ }\href {http://archive.org/details/in.ernet.dli.2015.223699} {\emph {\bibinfo {title} {Mathematical {Methods} {Of} {Statistics}}}}\ (\bibinfo  {publisher} {Princeton University Press},\ \bibinfo {year} {1946})\BibitemShut {NoStop}%
\bibitem [{\citenamefont {Rao}(1947)}]{rao_minimum_1947}%
  \BibitemOpen
  \bibfield  {author} {\bibinfo {author} {\bibfnamefont {C.~R.}\ \bibnamefont {Rao}},\ }\bibfield  {title} {\bibinfo {title} {Minimum variance and the estimation of several parameters},\ }\href {https://doi.org/10.1017/S0305004100023471} {\bibfield  {journal} {\bibinfo  {journal} {Mathematical Proceedings of the Cambridge Philosophical Society}\ }\textbf {\bibinfo {volume} {43}},\ \bibinfo {pages} {280} (\bibinfo {year} {1947})}\BibitemShut {NoStop}%
\bibitem [{\citenamefont {Hayashi}(2005)}]{hayashi_asymptotic_2005}%
  \BibitemOpen
  \bibfield  {author} {\bibinfo {author} {\bibfnamefont {M.}~\bibnamefont {Hayashi}},\ }\href {https://doi.org/10.1142/5630} {\emph {\bibinfo {title} {Asymptotic {Theory} of {Quantum} {Statistical} {Inference}: {Selected} {Papers}}}}\ (\bibinfo  {publisher} {World Scientific},\ \bibinfo {year} {2005})\BibitemShut {NoStop}%
\bibitem [{\citenamefont {Matsumoto}(2002)}]{matsumoto_new_2002}%
  \BibitemOpen
  \bibfield  {author} {\bibinfo {author} {\bibfnamefont {K.}~\bibnamefont {Matsumoto}},\ }\bibfield  {title} {\bibinfo {title} {A new approach to the {Cram{\'e}r}-{Rao}-type bound of the pure-state model},\ }\href {https://doi.org/10.1088/0305-4470/35/13/307} {\bibfield  {journal} {\bibinfo  {journal} {Journal of Physics A: Mathematical and General}\ }\textbf {\bibinfo {volume} {35}},\ \bibinfo {pages} {3111} (\bibinfo {year} {2002})}\BibitemShut {NoStop}%
\bibitem [{\citenamefont {Ragy}\ \emph {et~al.}(2016)\citenamefont {Ragy}, \citenamefont {Jarzyna},\ and\ \citenamefont {Demkowicz-Dobrza{\'n}ski}}]{ragy_compatibility_2016}%
  \BibitemOpen
  \bibfield  {author} {\bibinfo {author} {\bibfnamefont {S.}~\bibnamefont {Ragy}}, \bibinfo {author} {\bibfnamefont {M.}~\bibnamefont {Jarzyna}},\ and\ \bibinfo {author} {\bibfnamefont {R.}~\bibnamefont {Demkowicz-Dobrza{\'n}ski}},\ }\bibfield  {title} {\bibinfo {title} {Compatibility in multiparameter quantum metrology},\ }\href {https://doi.org/10.1103/PhysRevA.94.052108} {\bibfield  {journal} {\bibinfo  {journal} {Physical Review A}\ }\textbf {\bibinfo {volume} {94}},\ \bibinfo {pages} {052108} (\bibinfo {year} {2016})}\BibitemShut {NoStop}%
\bibitem [{Note1()}]{Note1}%
  \BibitemOpen
  \bibinfo {note} {Any projective measurement on a qubit system will produce a singular classical Fisher information matrix, because there are two dependent outcomes but more than one parameter~\cite {candeloro_dimension_2024}.}\BibitemShut {Stop}%
\bibitem [{\citenamefont {Matsumoto}(2005)}]{hayashi_geometrical_2005}%
  \BibitemOpen
  \bibfield  {author} {\bibinfo {author} {\bibfnamefont {K.}~\bibnamefont {Matsumoto}},\ }\bibfield  {title} {\bibinfo {title} {A {Geometrical} {Approach} to {Quantum} {Estimation} {Theory}},\ }in\ \href {https://doi.org/10.1142/9789812563071_0021} {\emph {\bibinfo {booktitle} {Asymptotic {Theory} of {Quantum} {Statistical} {Inference}}}}\ (\bibinfo  {publisher} {WORLD SCIENTIFIC},\ \bibinfo {year} {2005})\ pp.\ \bibinfo {pages} {305--350}\BibitemShut {NoStop}%
\bibitem [{\citenamefont {Huelga}\ \emph {et~al.}(1997)\citenamefont {Huelga}, \citenamefont {Macchiavello}, \citenamefont {Pellizzari}, \citenamefont {Ekert}, \citenamefont {Plenio},\ and\ \citenamefont {Cirac}}]{huelga_improvement_1997}%
  \BibitemOpen
  \bibfield  {author} {\bibinfo {author} {\bibfnamefont {S.~F.}\ \bibnamefont {Huelga}}, \bibinfo {author} {\bibfnamefont {C.}~\bibnamefont {Macchiavello}}, \bibinfo {author} {\bibfnamefont {T.}~\bibnamefont {Pellizzari}}, \bibinfo {author} {\bibfnamefont {A.~K.}\ \bibnamefont {Ekert}}, \bibinfo {author} {\bibfnamefont {M.~B.}\ \bibnamefont {Plenio}},\ and\ \bibinfo {author} {\bibfnamefont {J.~I.}\ \bibnamefont {Cirac}},\ }\bibfield  {title} {\bibinfo {title} {Improvement of {Frequency} {Standards} with {Quantum} {Entanglement}},\ }\href {https://doi.org/10.1103/PhysRevLett.79.3865} {\bibfield  {journal} {\bibinfo  {journal} {Physical Review Letters}\ }\textbf {\bibinfo {volume} {79}},\ \bibinfo {pages} {3865} (\bibinfo {year} {1997})}\BibitemShut {NoStop}%
\bibitem [{\citenamefont {Genoni}\ \emph {et~al.}(2011)\citenamefont {Genoni}, \citenamefont {Olivares},\ and\ \citenamefont {Paris}}]{genoni_optical_2011}%
  \BibitemOpen
  \bibfield  {author} {\bibinfo {author} {\bibfnamefont {M.~G.}\ \bibnamefont {Genoni}}, \bibinfo {author} {\bibfnamefont {S.}~\bibnamefont {Olivares}},\ and\ \bibinfo {author} {\bibfnamefont {M.~G.~A.}\ \bibnamefont {Paris}},\ }\bibfield  {title} {\bibinfo {title} {Optical {Phase} {Estimation} in the {Presence} of {Phase} {Diffusion}},\ }\href {https://doi.org/10.1103/PhysRevLett.106.153603} {\bibfield  {journal} {\bibinfo  {journal} {Physical Review Letters}\ }\textbf {\bibinfo {volume} {106}},\ \bibinfo {pages} {153603} (\bibinfo {year} {2011})}\BibitemShut {NoStop}%
\bibitem [{\citenamefont {Genoni}\ \emph {et~al.}(2012)\citenamefont {Genoni}, \citenamefont {Olivares}, \citenamefont {Brivio}, \citenamefont {Cialdi}, \citenamefont {Cipriani}, \citenamefont {Santamato}, \citenamefont {Vezzoli},\ and\ \citenamefont {Paris}}]{genoni_optical_2012}%
  \BibitemOpen
  \bibfield  {author} {\bibinfo {author} {\bibfnamefont {M.~G.}\ \bibnamefont {Genoni}}, \bibinfo {author} {\bibfnamefont {S.}~\bibnamefont {Olivares}}, \bibinfo {author} {\bibfnamefont {D.}~\bibnamefont {Brivio}}, \bibinfo {author} {\bibfnamefont {S.}~\bibnamefont {Cialdi}}, \bibinfo {author} {\bibfnamefont {D.}~\bibnamefont {Cipriani}}, \bibinfo {author} {\bibfnamefont {A.}~\bibnamefont {Santamato}}, \bibinfo {author} {\bibfnamefont {S.}~\bibnamefont {Vezzoli}},\ and\ \bibinfo {author} {\bibfnamefont {M.~G.~A.}\ \bibnamefont {Paris}},\ }\bibfield  {title} {\bibinfo {title} {Optical interferometry in the presence of large phase diffusion},\ }\href {https://doi.org/10.1103/PhysRevA.85.043817} {\bibfield  {journal} {\bibinfo  {journal} {Physical Review A}\ }\textbf {\bibinfo {volume} {85}},\ \bibinfo {pages} {043817} (\bibinfo {year} {2012})}\BibitemShut {NoStop}%
\bibitem [{\citenamefont {Vidrighin}\ \emph {et~al.}(2014)\citenamefont {Vidrighin}, \citenamefont {Donati}, \citenamefont {Genoni}, \citenamefont {Jin}, \citenamefont {Kolthammer}, \citenamefont {Kim}, \citenamefont {Datta}, \citenamefont {Barbieri},\ and\ \citenamefont {Walmsley}}]{vidrighin_joint_2014}%
  \BibitemOpen
  \bibfield  {author} {\bibinfo {author} {\bibfnamefont {M.~D.}\ \bibnamefont {Vidrighin}}, \bibinfo {author} {\bibfnamefont {G.}~\bibnamefont {Donati}}, \bibinfo {author} {\bibfnamefont {M.~G.}\ \bibnamefont {Genoni}}, \bibinfo {author} {\bibfnamefont {X.-M.}\ \bibnamefont {Jin}}, \bibinfo {author} {\bibfnamefont {W.~S.}\ \bibnamefont {Kolthammer}}, \bibinfo {author} {\bibfnamefont {M.~S.}\ \bibnamefont {Kim}}, \bibinfo {author} {\bibfnamefont {A.}~\bibnamefont {Datta}}, \bibinfo {author} {\bibfnamefont {M.}~\bibnamefont {Barbieri}},\ and\ \bibinfo {author} {\bibfnamefont {I.~A.}\ \bibnamefont {Walmsley}},\ }\bibfield  {title} {\bibinfo {title} {Joint estimation of phase and phase diffusion for quantum metrology},\ }\href {https://doi.org/10.1038/ncomms4532} {\bibfield  {journal} {\bibinfo  {journal} {Nature Communications}\ }\textbf {\bibinfo {volume} {5}},\ \bibinfo {pages} {3532} (\bibinfo {year} {2014})}\BibitemShut {NoStop}%
\bibitem [{\citenamefont {Yuen}\ and\ \citenamefont {Lax}(1973)}]{yuen_multiple-parameter_1973}%
  \BibitemOpen
  \bibfield  {author} {\bibinfo {author} {\bibfnamefont {H.}~\bibnamefont {Yuen}}\ and\ \bibinfo {author} {\bibfnamefont {M.}~\bibnamefont {Lax}},\ }\bibfield  {title} {\bibinfo {title} {Multiple-parameter quantum estimation and measurement of nonselfadjoint observables},\ }\href {https://doi.org/10.1109/TIT.1973.1055103} {\bibfield  {journal} {\bibinfo  {journal} {IEEE Transactions on Information Theory}\ }\textbf {\bibinfo {volume} {19}},\ \bibinfo {pages} {740} (\bibinfo {year} {1973})}\BibitemShut {NoStop}%
\bibitem [{\citenamefont {Gill}\ and\ \citenamefont {Massar}(2000)}]{gill_state_2000}%
  \BibitemOpen
  \bibfield  {author} {\bibinfo {author} {\bibfnamefont {R.~D.}\ \bibnamefont {Gill}}\ and\ \bibinfo {author} {\bibfnamefont {S.}~\bibnamefont {Massar}},\ }\bibfield  {title} {\bibinfo {title} {State estimation for large ensembles},\ }\href {https://doi.org/10.1103/PhysRevA.61.042312} {\bibfield  {journal} {\bibinfo  {journal} {Physical Review A}\ }\textbf {\bibinfo {volume} {61}},\ \bibinfo {pages} {042312} (\bibinfo {year} {2000})}\BibitemShut {NoStop}%
\bibitem [{\citenamefont {Holevo}(1973)}]{holevo_statistical_1973}%
  \BibitemOpen
  \bibfield  {author} {\bibinfo {author} {\bibfnamefont {A.~S.}\ \bibnamefont {Holevo}},\ }\bibfield  {title} {\bibinfo {title} {Statistical decision theory for quantum systems},\ }\href {https://doi.org/10.1016/0047-259X(73)90028-6} {\bibfield  {journal} {\bibinfo  {journal} {Journal of Multivariate Analysis}\ }\textbf {\bibinfo {volume} {3}},\ \bibinfo {pages} {337} (\bibinfo {year} {1973})}\BibitemShut {NoStop}%
\bibitem [{\citenamefont {Holevo}(2011)}]{holevo_probabilistic_2011}%
  \BibitemOpen
  \bibfield  {author} {\bibinfo {author} {\bibfnamefont {A.~S.}\ \bibnamefont {Holevo}},\ }\href@noop {} {\emph {\bibinfo {title} {Probabilistic and statistical aspects of quantum theory}}},\ \bibinfo {series} {Quaderni monographs}\ No.~\bibinfo {number} {1}\ (\bibinfo  {publisher} {Ed. della Normale},\ \bibinfo {address} {Pisa},\ \bibinfo {year} {2011})\BibitemShut {NoStop}%
\bibitem [{\citenamefont {Nagaoka}(2005{\natexlab{a}})}]{nagaoka_new_2005}%
  \BibitemOpen
  \bibfield  {author} {\bibinfo {author} {\bibfnamefont {H.}~\bibnamefont {Nagaoka}},\ }\bibfield  {title} {\bibinfo {title} {A {New} {Approach} to {Cram{\'e}r}-{Rao} {Bounds} for {Quantum} {State} {Estimation}},\ }in\ \href {https://doi.org/10.1142/9789812563071_0009} {\emph {\bibinfo {booktitle} {Asymptotic {Theory} of {Quantum} {Statistical} {Inference}}}}\ (\bibinfo  {publisher} {WORLD SCIENTIFIC},\ \bibinfo {year} {2005})\ pp.\ \bibinfo {pages} {100--112}\BibitemShut {NoStop}%
\bibitem [{\citenamefont {Nagaoka}(2005{\natexlab{b}})}]{nagaoka_generalization_2005}%
  \BibitemOpen
  \bibfield  {author} {\bibinfo {author} {\bibfnamefont {H.}~\bibnamefont {Nagaoka}},\ }\bibfield  {title} {\bibinfo {title} {A {Generalization} of the {Simultaneous} {Diagonalization} of {Hermitian} {Matrices} and its {Relation} to {Quantum} {Estimation} {Theory}},\ }in\ \href {https://doi.org/10.1142/9789812563071_0012} {\emph {\bibinfo {booktitle} {Asymptotic {Theory} of {Quantum} {Statistical} {Inference}}}}\ (\bibinfo  {publisher} {WORLD SCIENTIFIC},\ \bibinfo {year} {2005})\ pp.\ \bibinfo {pages} {133--149}\BibitemShut {NoStop}%
\bibitem [{\citenamefont {Conlon}\ \emph {et~al.}(2021)\citenamefont {Conlon}, \citenamefont {Suzuki}, \citenamefont {Lam},\ and\ \citenamefont {Assad}}]{conlon_efficient_2021}%
  \BibitemOpen
  \bibfield  {author} {\bibinfo {author} {\bibfnamefont {L.~O.}\ \bibnamefont {Conlon}}, \bibinfo {author} {\bibfnamefont {J.}~\bibnamefont {Suzuki}}, \bibinfo {author} {\bibfnamefont {P.~K.}\ \bibnamefont {Lam}},\ and\ \bibinfo {author} {\bibfnamefont {S.~M.}\ \bibnamefont {Assad}},\ }\bibfield  {title} {\bibinfo {title} {Efficient computation of the {Nagaoka}{\textendash}{Hayashi} bound for multiparameter estimation with separable measurements},\ }\href {https://doi.org/10.1038/s41534-021-00414-1} {\bibfield  {journal} {\bibinfo  {journal} {npj Quantum Information}\ }\textbf {\bibinfo {volume} {7}},\ \bibinfo {pages} {1} (\bibinfo {year} {2021})}\BibitemShut {NoStop}%
\bibitem [{\citenamefont {Lu}\ and\ \citenamefont {Wang}(2021)}]{lu_incorporating_2021}%
  \BibitemOpen
  \bibfield  {author} {\bibinfo {author} {\bibfnamefont {X.-M.}\ \bibnamefont {Lu}}\ and\ \bibinfo {author} {\bibfnamefont {X.}~\bibnamefont {Wang}},\ }\bibfield  {title} {\bibinfo {title} {Incorporating {Heisenberg}'s {Uncertainty} {Principle} into {Quantum} {Multiparameter} {Estimation}},\ }\href {https://doi.org/10.1103/PhysRevLett.126.120503} {\bibfield  {journal} {\bibinfo  {journal} {Physical Review Letters}\ }\textbf {\bibinfo {volume} {126}},\ \bibinfo {pages} {120503} (\bibinfo {year} {2021})}\BibitemShut {NoStop}%
\bibitem [{\citenamefont {Yung}\ \emph {et~al.}(2024)\citenamefont {Yung}, \citenamefont {Conlon}, \citenamefont {Zhao}, \citenamefont {Lam},\ and\ \citenamefont {Assad}}]{yung_comparison_2024}%
  \BibitemOpen
  \bibfield  {author} {\bibinfo {author} {\bibfnamefont {S.~K.}\ \bibnamefont {Yung}}, \bibinfo {author} {\bibfnamefont {L.~O.}\ \bibnamefont {Conlon}}, \bibinfo {author} {\bibfnamefont {J.}~\bibnamefont {Zhao}}, \bibinfo {author} {\bibfnamefont {P.~K.}\ \bibnamefont {Lam}},\ and\ \bibinfo {author} {\bibfnamefont {S.~M.}\ \bibnamefont {Assad}},\ }\bibfield  {title} {\bibinfo {title} {Comparison of estimation limits for quantum two-parameter estimation},\ }\href {https://doi.org/10.1103/PhysRevResearch.6.033315} {\bibfield  {journal} {\bibinfo  {journal} {Physical Review Research}\ }\textbf {\bibinfo {volume} {6}},\ \bibinfo {pages} {033315} (\bibinfo {year} {2024})}\BibitemShut {NoStop}%
\bibitem [{Note2()}]{Note2}%
  \BibitemOpen
  \bibinfo {note} {See Supplemental Material for additional details on the phase--dephasing estimation problem and the Fock state displacement sensing problem, details on the Quantinuum experiment, and the proof of Theorem 1.}\BibitemShut {Stop}%
\bibitem [{Note3()}]{Note3}%
  \BibitemOpen
  \bibinfo {note} {Quantinuum H1-1, \protect \url {https://quantinuum.com/}, accessed April 10--11 2025.}\BibitemShut {Stop}%
\bibitem [{\citenamefont {Suzuki}\ \emph {et~al.}(2020)\citenamefont {Suzuki}, \citenamefont {Yang},\ and\ \citenamefont {Hayashi}}]{suzuki_quantum_2020}%
  \BibitemOpen
  \bibfield  {author} {\bibinfo {author} {\bibfnamefont {J.}~\bibnamefont {Suzuki}}, \bibinfo {author} {\bibfnamefont {Y.}~\bibnamefont {Yang}},\ and\ \bibinfo {author} {\bibfnamefont {M.}~\bibnamefont {Hayashi}},\ }\bibfield  {title} {\bibinfo {title} {Quantum state estimation with nuisance parameters},\ }\href {https://doi.org/10.1088/1751-8121/ab8b78} {\bibfield  {journal} {\bibinfo  {journal} {Journal of Physics A: Mathematical and Theoretical}\ }\textbf {\bibinfo {volume} {53}},\ \bibinfo {pages} {453001} (\bibinfo {year} {2020})}\BibitemShut {NoStop}%
\bibitem [{\citenamefont {Arvidsson-Shukur}\ \emph {et~al.}(2020)\citenamefont {Arvidsson-Shukur}, \citenamefont {Yunger~Halpern}, \citenamefont {Lepage}, \citenamefont {Lasek}, \citenamefont {Barnes},\ and\ \citenamefont {Lloyd}}]{arvidsson-shukur_quantum_2020}%
  \BibitemOpen
  \bibfield  {author} {\bibinfo {author} {\bibfnamefont {D.~R.~M.}\ \bibnamefont {Arvidsson-Shukur}}, \bibinfo {author} {\bibfnamefont {N.}~\bibnamefont {Yunger~Halpern}}, \bibinfo {author} {\bibfnamefont {H.~V.}\ \bibnamefont {Lepage}}, \bibinfo {author} {\bibfnamefont {A.~A.}\ \bibnamefont {Lasek}}, \bibinfo {author} {\bibfnamefont {C.~H.~W.}\ \bibnamefont {Barnes}},\ and\ \bibinfo {author} {\bibfnamefont {S.}~\bibnamefont {Lloyd}},\ }\bibfield  {title} {\bibinfo {title} {Quantum advantage in postselected metrology},\ }\href {https://doi.org/10.1038/s41467-020-17559-w} {\bibfield  {journal} {\bibinfo  {journal} {Nature Communications}\ }\textbf {\bibinfo {volume} {11}},\ \bibinfo {pages} {3775} (\bibinfo {year} {2020})}\BibitemShut {NoStop}%
\bibitem [{\citenamefont {Das}\ \emph {et~al.}(2023)\citenamefont {Das}, \citenamefont {Modak},\ and\ \citenamefont {Bera}}]{das_saturating_2023}%
  \BibitemOpen
  \bibfield  {author} {\bibinfo {author} {\bibfnamefont {S.}~\bibnamefont {Das}}, \bibinfo {author} {\bibfnamefont {S.}~\bibnamefont {Modak}},\ and\ \bibinfo {author} {\bibfnamefont {M.~N.}\ \bibnamefont {Bera}},\ }\bibfield  {title} {\bibinfo {title} {Saturating quantum advantages in postselected metrology with the positive {Kirkwood}-{Dirac} distribution},\ }\href {https://doi.org/10.1103/PhysRevA.107.042413} {\bibfield  {journal} {\bibinfo  {journal} {Physical Review A}\ }\textbf {\bibinfo {volume} {107}},\ \bibinfo {pages} {042413} (\bibinfo {year} {2023})}\BibitemShut {NoStop}%
\bibitem [{\citenamefont {Assad}\ \emph {et~al.}(2020)\citenamefont {Assad}, \citenamefont {Li}, \citenamefont {Liu}, \citenamefont {Zhao}, \citenamefont {Zhao}, \citenamefont {Lam}, \citenamefont {Ou},\ and\ \citenamefont {Li}}]{assad_accessible_2020}%
  \BibitemOpen
  \bibfield  {author} {\bibinfo {author} {\bibfnamefont {S.~M.}\ \bibnamefont {Assad}}, \bibinfo {author} {\bibfnamefont {J.}~\bibnamefont {Li}}, \bibinfo {author} {\bibfnamefont {Y.}~\bibnamefont {Liu}}, \bibinfo {author} {\bibfnamefont {N.}~\bibnamefont {Zhao}}, \bibinfo {author} {\bibfnamefont {W.}~\bibnamefont {Zhao}}, \bibinfo {author} {\bibfnamefont {P.~K.}\ \bibnamefont {Lam}}, \bibinfo {author} {\bibfnamefont {Z.~Y.}\ \bibnamefont {Ou}},\ and\ \bibinfo {author} {\bibfnamefont {X.}~\bibnamefont {Li}},\ }\bibfield  {title} {\bibinfo {title} {Accessible precisions for estimating two conjugate parameters using {Gaussian} probes},\ }\href {https://doi.org/10.1103/PhysRevResearch.2.023182} {\bibfield  {journal} {\bibinfo  {journal} {Physical Review Research}\ }\textbf {\bibinfo {volume} {2}},\ \bibinfo {pages} {023182} (\bibinfo {year} {2020})}\BibitemShut {NoStop}%
\bibitem [{\citenamefont {Bradshaw}\ \emph {et~al.}(2017)\citenamefont {Bradshaw}, \citenamefont {Assad},\ and\ \citenamefont {Lam}}]{bradshaw_tight_2017}%
  \BibitemOpen
  \bibfield  {author} {\bibinfo {author} {\bibfnamefont {M.}~\bibnamefont {Bradshaw}}, \bibinfo {author} {\bibfnamefont {S.~M.}\ \bibnamefont {Assad}},\ and\ \bibinfo {author} {\bibfnamefont {P.~K.}\ \bibnamefont {Lam}},\ }\bibfield  {title} {\bibinfo {title} {A tight {Cram{\'e}r}{\textendash}{Rao} bound for joint parameter estimation with a pure two-mode squeezed probe},\ }\href {https://doi.org/10.1016/j.physleta.2017.06.024} {\bibfield  {journal} {\bibinfo  {journal} {Physics Letters A}\ }\textbf {\bibinfo {volume} {381}},\ \bibinfo {pages} {2598} (\bibinfo {year} {2017})}\BibitemShut {NoStop}%
\bibitem [{\citenamefont {Bradshaw}\ \emph {et~al.}(2018)\citenamefont {Bradshaw}, \citenamefont {Lam},\ and\ \citenamefont {Assad}}]{bradshaw_ultimate_2018}%
  \BibitemOpen
  \bibfield  {author} {\bibinfo {author} {\bibfnamefont {M.}~\bibnamefont {Bradshaw}}, \bibinfo {author} {\bibfnamefont {P.~K.}\ \bibnamefont {Lam}},\ and\ \bibinfo {author} {\bibfnamefont {S.~M.}\ \bibnamefont {Assad}},\ }\bibfield  {title} {\bibinfo {title} {Ultimate precision of joint quadrature parameter estimation with a {Gaussian} probe},\ }\href {https://doi.org/10.1103/PhysRevA.97.012106} {\bibfield  {journal} {\bibinfo  {journal} {Physical Review A}\ }\textbf {\bibinfo {volume} {97}},\ \bibinfo {pages} {012106} (\bibinfo {year} {2018})}\BibitemShut {NoStop}%
\bibitem [{\citenamefont {Gardner}\ \emph {et~al.}(2024)\citenamefont {Gardner}, \citenamefont {Gefen}, \citenamefont {Haine}, \citenamefont {Hope},\ and\ \citenamefont {Chen}}]{gardner_achieving_2024}%
  \BibitemOpen
  \bibfield  {author} {\bibinfo {author} {\bibfnamefont {J.~W.}\ \bibnamefont {Gardner}}, \bibinfo {author} {\bibfnamefont {T.}~\bibnamefont {Gefen}}, \bibinfo {author} {\bibfnamefont {S.~A.}\ \bibnamefont {Haine}}, \bibinfo {author} {\bibfnamefont {J.~J.}\ \bibnamefont {Hope}},\ and\ \bibinfo {author} {\bibfnamefont {Y.}~\bibnamefont {Chen}},\ }\bibfield  {title} {\bibinfo {title} {Achieving the {Fundamental} {Quantum} {Limit} of {Linear} {Waveform} {Estimation}},\ }\href {https://doi.org/10.1103/PhysRevLett.132.130801} {\bibfield  {journal} {\bibinfo  {journal} {Physical Review Letters}\ }\textbf {\bibinfo {volume} {132}},\ \bibinfo {pages} {130801} (\bibinfo {year} {2024})}\BibitemShut {NoStop}%
\bibitem [{\citenamefont {Li}\ and\ \citenamefont {Lu}(2024)}]{li_general_2024}%
  \BibitemOpen
  \bibfield  {author} {\bibinfo {author} {\bibfnamefont {G.}~\bibnamefont {Li}}\ and\ \bibinfo {author} {\bibfnamefont {X.-M.}\ \bibnamefont {Lu}},\ }\href {https://doi.org/10.48550/arXiv.2412.15031} {\bibinfo {title} {General tradeoff relation of fundamental quantum limits for linear multiparameter estimation}} (\bibinfo {year} {2024}),\ \bibinfo {note} {arXiv:2412.15031 [quant-ph]}\BibitemShut {NoStop}%
\bibitem [{\citenamefont {Xia}\ \emph {et~al.}(2023)\citenamefont {Xia}, \citenamefont {Huang}, \citenamefont {Li}, \citenamefont {Wang},\ and\ \citenamefont {Zeng}}]{xia_toward_2023}%
  \BibitemOpen
  \bibfield  {author} {\bibinfo {author} {\bibfnamefont {B.}~\bibnamefont {Xia}}, \bibinfo {author} {\bibfnamefont {J.}~\bibnamefont {Huang}}, \bibinfo {author} {\bibfnamefont {H.}~\bibnamefont {Li}}, \bibinfo {author} {\bibfnamefont {H.}~\bibnamefont {Wang}},\ and\ \bibinfo {author} {\bibfnamefont {G.}~\bibnamefont {Zeng}},\ }\bibfield  {title} {\bibinfo {title} {Toward incompatible quantum limits on multiparameter estimation},\ }\href {https://doi.org/10.1038/s41467-023-36661-3} {\bibfield  {journal} {\bibinfo  {journal} {Nature Communications}\ }\textbf {\bibinfo {volume} {14}},\ \bibinfo {pages} {1021} (\bibinfo {year} {2023})}\BibitemShut {NoStop}%
\bibitem [{\citenamefont {Crowley}\ \emph {et~al.}(2014)\citenamefont {Crowley}, \citenamefont {Datta}, \citenamefont {Barbieri},\ and\ \citenamefont {Walmsley}}]{crowley_tradeoff_2014}%
  \BibitemOpen
  \bibfield  {author} {\bibinfo {author} {\bibfnamefont {P.~J.~D.}\ \bibnamefont {Crowley}}, \bibinfo {author} {\bibfnamefont {A.}~\bibnamefont {Datta}}, \bibinfo {author} {\bibfnamefont {M.}~\bibnamefont {Barbieri}},\ and\ \bibinfo {author} {\bibfnamefont {I.~A.}\ \bibnamefont {Walmsley}},\ }\bibfield  {title} {\bibinfo {title} {Tradeoff in simultaneous quantum-limited phase and loss estimation in interferometry},\ }\href {https://doi.org/10.1103/PhysRevA.89.023845} {\bibfield  {journal} {\bibinfo  {journal} {Physical Review A}\ }\textbf {\bibinfo {volume} {89}},\ \bibinfo {pages} {023845} (\bibinfo {year} {2014})}\BibitemShut {NoStop}%
\bibitem [{\citenamefont {Suzuki}(2020)}]{suzuki_nuisance_2020}%
  \BibitemOpen
  \bibfield  {author} {\bibinfo {author} {\bibfnamefont {J.}~\bibnamefont {Suzuki}},\ }\bibfield  {title} {\bibinfo {title} {Nuisance parameter problem in quantum estimation theory: tradeoff relation and qubit examples},\ }\href {https://doi.org/10.1088/1751-8121/ab8672} {\bibfield  {journal} {\bibinfo  {journal} {Journal of Physics A: Mathematical and Theoretical}\ }\textbf {\bibinfo {volume} {53}},\ \bibinfo {pages} {264001} (\bibinfo {year} {2020})}\BibitemShut {NoStop}%
\bibitem [{\citenamefont {Hou}\ \emph {et~al.}(2018)\citenamefont {Hou}, \citenamefont {Tang}, \citenamefont {Shang}, \citenamefont {Zhu}, \citenamefont {Li}, \citenamefont {Yuan}, \citenamefont {Wu}, \citenamefont {Xiang}, \citenamefont {Li},\ and\ \citenamefont {Guo}}]{hou_deterministic_2018}%
  \BibitemOpen
  \bibfield  {author} {\bibinfo {author} {\bibfnamefont {Z.}~\bibnamefont {Hou}}, \bibinfo {author} {\bibfnamefont {J.-F.}\ \bibnamefont {Tang}}, \bibinfo {author} {\bibfnamefont {J.}~\bibnamefont {Shang}}, \bibinfo {author} {\bibfnamefont {H.}~\bibnamefont {Zhu}}, \bibinfo {author} {\bibfnamefont {J.}~\bibnamefont {Li}}, \bibinfo {author} {\bibfnamefont {Y.}~\bibnamefont {Yuan}}, \bibinfo {author} {\bibfnamefont {K.-D.}\ \bibnamefont {Wu}}, \bibinfo {author} {\bibfnamefont {G.-Y.}\ \bibnamefont {Xiang}}, \bibinfo {author} {\bibfnamefont {C.-F.}\ \bibnamefont {Li}},\ and\ \bibinfo {author} {\bibfnamefont {G.-C.}\ \bibnamefont {Guo}},\ }\bibfield  {title} {\bibinfo {title} {Deterministic realization of collective measurements via photonic quantum walks},\ }\href {https://doi.org/10.1038/s41467-018-03849-x} {\bibfield  {journal} {\bibinfo  {journal} {Nature Communications}\ }\textbf {\bibinfo {volume} {9}},\ \bibinfo {pages} {1414} (\bibinfo {year} {2018})}\BibitemShut {NoStop}%
\bibitem [{\citenamefont {Conlon}\ \emph {et~al.}(2023{\natexlab{a}})\citenamefont {Conlon}, \citenamefont {Vogl}, \citenamefont {Marciniak}, \citenamefont {Pogorelov}, \citenamefont {Yung}, \citenamefont {Eilenberger}, \citenamefont {Berry}, \citenamefont {Santana}, \citenamefont {Blatt}, \citenamefont {Monz}, \citenamefont {Lam},\ and\ \citenamefont {Assad}}]{conlon_approaching_2023}%
  \BibitemOpen
  \bibfield  {author} {\bibinfo {author} {\bibfnamefont {L.~O.}\ \bibnamefont {Conlon}}, \bibinfo {author} {\bibfnamefont {T.}~\bibnamefont {Vogl}}, \bibinfo {author} {\bibfnamefont {C.~D.}\ \bibnamefont {Marciniak}}, \bibinfo {author} {\bibfnamefont {I.}~\bibnamefont {Pogorelov}}, \bibinfo {author} {\bibfnamefont {S.~K.}\ \bibnamefont {Yung}}, \bibinfo {author} {\bibfnamefont {F.}~\bibnamefont {Eilenberger}}, \bibinfo {author} {\bibfnamefont {D.~W.}\ \bibnamefont {Berry}}, \bibinfo {author} {\bibfnamefont {F.~S.}\ \bibnamefont {Santana}}, \bibinfo {author} {\bibfnamefont {R.}~\bibnamefont {Blatt}}, \bibinfo {author} {\bibfnamefont {T.}~\bibnamefont {Monz}}, \bibinfo {author} {\bibfnamefont {P.~K.}\ \bibnamefont {Lam}},\ and\ \bibinfo {author} {\bibfnamefont {S.~M.}\ \bibnamefont {Assad}},\ }\bibfield  {title} {\bibinfo {title} {Approaching optimal entangling collective measurements on quantum computing platforms},\ }\href {https://doi.org/10.1038/s41567-022-01875-7} {\bibfield  {journal} {\bibinfo  {journal} {Nature Physics}\ }\textbf {\bibinfo {volume} {19}},\ \bibinfo {pages} {351} (\bibinfo {year} {2023}{\natexlab{a}})}\BibitemShut {NoStop}%
\bibitem [{\citenamefont {Conlon}\ \emph {et~al.}(2023{\natexlab{b}})\citenamefont {Conlon}, \citenamefont {Eilenberger}, \citenamefont {Lam},\ and\ \citenamefont {Assad}}]{conlon_discriminating_2023}%
  \BibitemOpen
  \bibfield  {author} {\bibinfo {author} {\bibfnamefont {L.~O.}\ \bibnamefont {Conlon}}, \bibinfo {author} {\bibfnamefont {F.}~\bibnamefont {Eilenberger}}, \bibinfo {author} {\bibfnamefont {P.~K.}\ \bibnamefont {Lam}},\ and\ \bibinfo {author} {\bibfnamefont {S.~M.}\ \bibnamefont {Assad}},\ }\bibfield  {title} {\bibinfo {title} {Discriminating mixed qubit states with collective measurements},\ }\href {https://doi.org/10.1038/s42005-023-01454-z} {\bibfield  {journal} {\bibinfo  {journal} {Communications Physics}\ }\textbf {\bibinfo {volume} {6}},\ \bibinfo {pages} {1} (\bibinfo {year} {2023}{\natexlab{b}})}\BibitemShut {NoStop}%
\bibitem [{\citenamefont {Zhou}\ \emph {et~al.}(2025)\citenamefont {Zhou}, \citenamefont {Yi}, \citenamefont {Yan}, \citenamefont {Hou}, \citenamefont {Zhu}, \citenamefont {Xiang}, \citenamefont {Li},\ and\ \citenamefont {Guo}}]{zhou_experimental_2025}%
  \BibitemOpen
  \bibfield  {author} {\bibinfo {author} {\bibfnamefont {K.}~\bibnamefont {Zhou}}, \bibinfo {author} {\bibfnamefont {C.}~\bibnamefont {Yi}}, \bibinfo {author} {\bibfnamefont {W.-Z.}\ \bibnamefont {Yan}}, \bibinfo {author} {\bibfnamefont {Z.}~\bibnamefont {Hou}}, \bibinfo {author} {\bibfnamefont {H.}~\bibnamefont {Zhu}}, \bibinfo {author} {\bibfnamefont {G.-Y.}\ \bibnamefont {Xiang}}, \bibinfo {author} {\bibfnamefont {C.-F.}\ \bibnamefont {Li}},\ and\ \bibinfo {author} {\bibfnamefont {G.-C.}\ \bibnamefont {Guo}},\ }\bibfield  {title} {\bibinfo {title} {Experimental {Realization} of {Genuine} {Three}-{Copy} {Collective} {Measurements} for {Optimal} {Information} {Extraction}},\ }\href {https://doi.org/10.1103/PhysRevLett.134.210201} {\bibfield  {journal} {\bibinfo  {journal} {Physical Review Letters}\ }\textbf {\bibinfo {volume} {134}},\ \bibinfo {pages} {210201} (\bibinfo {year} {2025})}\BibitemShut {NoStop}%
\bibitem [{\citenamefont {Massar}\ and\ \citenamefont {Popescu}(1995)}]{massar_optimal_1995}%
  \BibitemOpen
  \bibfield  {author} {\bibinfo {author} {\bibfnamefont {S.}~\bibnamefont {Massar}}\ and\ \bibinfo {author} {\bibfnamefont {S.}~\bibnamefont {Popescu}},\ }\bibfield  {title} {\bibinfo {title} {Optimal {Extraction} of {Information} from {Finite} {Quantum} {Ensembles}},\ }\href {https://doi.org/10.1103/PhysRevLett.74.1259} {\bibfield  {journal} {\bibinfo  {journal} {Physical Review Letters}\ }\textbf {\bibinfo {volume} {74}},\ \bibinfo {pages} {1259} (\bibinfo {year} {1995})}\BibitemShut {NoStop}%
\bibitem [{\citenamefont {Candeloro}\ \emph {et~al.}(2024)\citenamefont {Candeloro}, \citenamefont {Pazhotan},\ and\ \citenamefont {Paris}}]{candeloro_dimension_2024}%
  \BibitemOpen
  \bibfield  {author} {\bibinfo {author} {\bibfnamefont {A.}~\bibnamefont {Candeloro}}, \bibinfo {author} {\bibfnamefont {Z.}~\bibnamefont {Pazhotan}},\ and\ \bibinfo {author} {\bibfnamefont {M.~G.~A.}\ \bibnamefont {Paris}},\ }\bibfield  {title} {\bibinfo {title} {Dimension matters: precision and incompatibility in multi-parameter quantum estimation models},\ }\href {https://doi.org/10.1088/2058-9565/ad7498} {\bibfield  {journal} {\bibinfo  {journal} {Quantum Science and Technology}\ }\textbf {\bibinfo {volume} {9}},\ \bibinfo {pages} {045045} (\bibinfo {year} {2024})}\BibitemShut {NoStop}%
\end{thebibliography}%

\end{document}


\title{Supplemental Material for \\
\textit{Saturating the Quantum Cram\'er--Rao Bound in Prioritised Parameter Estimation}}
\author{Simon K. Yung}
\email{sksyung@gmail.com}
\affiliation{Centre for Quantum Computation and Communication Technology, Department of Quantum Science and Technology, Research School of Physics, The Australian National University, Canberra, ACT 2601, Australia.}
\author{Aritra Das}
\affiliation{Centre for Quantum Computation and Communication Technology, Department of Quantum Science and Technology, Research School of Physics, The Australian National University, Canberra, ACT 2601, Australia.}
\author{Jun Suzuki}
\affiliation{Graduate School of Informatics and Engineering, The University of Electro-Communications, Tokyo 182-8585, Japan.}

\author{Ping Koy Lam}
\affiliation{A*STAR Quantum Innovation Centre (Q.INC), Agency for Science, Technology and Research (A*STAR), 2 Fusionopolis Way, Innovis, 138634, Singapore.}
\affiliation{Centre for Quantum Computation and Communication Technology, Department of Quantum Science and Technology, Research School of Physics, The Australian National University, Canberra, ACT 2601, Australia.}
\affiliation{Centre for Quantum Technologies, National University of Singapore, 3 Science Drive 2, Singapore 117543, Singapore.}
\author{Jie Zhao}
\email{jie.zhao@anu.edu.au}
\affiliation{Centre for Quantum Computation and Communication Technology, Department of Quantum Science and Technology, Research School of Physics, The Australian National University, Canberra, ACT 2601, Australia.}
\author{Lorc\'an O. Conlon}
\altaffiliation{Present address: Joint Quantum Institute and Joint Center for Quantum Information and Computer Science, NIST/University of Maryland, College Park, Maryland 20742, USA.}
\affiliation{A*STAR Quantum Innovation Centre (Q.INC), Agency for Science, Technology and Research (A*STAR), 2 Fusionopolis Way, Innovis, 138634, Singapore.}
\affiliation{Centre for Quantum Technologies, National University of Singapore, 3 Science Drive 2, Singapore 117543, Singapore.}

\author{Syed M. Assad}
\email{cqtsma@gmail.com}
\affiliation{A*STAR Quantum Innovation Centre (Q.INC), Agency for Science, Technology and Research (A*STAR), 2 Fusionopolis Way, Innovis, 138634, Singapore.}

\date{\today}

\maketitle

\section{Phase--Dephasing estimation Details}
We consider a channel in which an initial probe state $(\ket{0}+\ket{1})(\bra{0}+\bra{1})/2$ undergoes a phase shift $\rho \rightarrow e^{-i\phi\sigma_z/2}\rho e^{i\phi\sigma_z/2}$ and dephasing $\rho \rightarrow (1-\Delta/2)\rho  + (\Delta/2) \sigma_z\rho\sigma_z$. The resultant density operator is
\begin{equation}
	\rho(\phi,\Delta) = \frac{1}{2}\begin{pmatrix}
		1 & (1-\Delta)e^{-i\phi} \\
		(1-\Delta)e^{i\phi} & 1 
	\end{pmatrix}.
\end{equation}
\subsection{Single-copy measurements}
The SLD operators with respect to $\phi$ and $\Delta$ are 
\begin{align}
	L_\phi &= \begin{pmatrix}
		0 & i(1-\Delta)e^{-i\phi} \\
		-i(1-\Delta)e^{i\phi} & 0 
	\end{pmatrix}, \\
	L_\Delta &= \frac{1}{\Delta(2-\Delta)}\begin{pmatrix}
		1-\Delta & -e^{-i\phi} \\
		-e^{i\phi} & 1-\Delta 
	\end{pmatrix},
\end{align}
and the quantum Fisher information is
\begin{equation}
	J = \begin{pmatrix}
		(1-\Delta)^2 & 0 \\
		0 & \frac{1}{\Delta(2-\Delta)}
	\end{pmatrix}.
\end{equation}
The eigensystems of the SLD operators are
\begin{align}
	L_\phi:& \begin{matrix}
		1-\Delta  & -(1-\Delta) \\
		\begin{pmatrix}
			-ie^{-i\phi} \\
			1
		\end{pmatrix} & \begin{pmatrix}
			ie^{-i\phi} \\
			1
		\end{pmatrix}
	\end{matrix}, \quad 
	L_\Delta: \begin{matrix}
		\frac{1}{\Delta} & \frac{-1}{2-\Delta} \\
		\begin{pmatrix}
			-e^{-i\phi} \\
			1
		\end{pmatrix} & \begin{pmatrix}
			e^{-i\phi} \\
			1
		\end{pmatrix}
	\end{matrix}.
\end{align}
The dependence of the eigensystems on the parameters says that the optimal measurements for estimating each parameter (individually) depend on the values of the parameters. For the $\phi$-optimal measurement, the optimal POVM does not depend on $\Delta$ but the estimator does, so knowledge of $\Delta$ is required to obtain an unbiased estimate of $\phi$. Explicitly, the estimator is
\begin{equation}
	\hat{\phi} = \phi_0+\frac{1}{1-\Delta}f_1 - \frac{1}{1-\Delta}f_2, \quad \Pi_1 = \frac{1}{2} \begin{pmatrix}
			-ie^{-i\phi_0} \\
			1
		\end{pmatrix}\begin{pmatrix}
			ie^{i\phi_0} &1
		\end{pmatrix}, \ \Pi_2 = \frac{1}{2}\begin{pmatrix}
			ie^{-i\phi} \\
			1
		\end{pmatrix}\begin{pmatrix}
			-ie^{i\phi} & 1
		\end{pmatrix}
\end{equation}
where $f_{1(2)}$ is the observed frequency of the POVM element $\Pi_{1(2)}$ which are the projectors onto the $L_\phi$ eigenspaces, and $\phi_0$ is the true value of $\phi$.  

Furthermore, a measurement in either eigenbasis gives no information about the other parameter. This is exemplified in the trade-off relation
\begin{equation}
	\left(V(\hat{\phi})-\underbrace{\frac{1}{(1-\Delta)^2}}_{[J^{-1}]_{\phi\phi}}\right)\left(V(\hat{\Delta})-\underbrace{\Delta(2-\Delta)}_{[J^{-1}]_{\Delta\Delta}}\right)\geq \underbrace{\frac{\Delta (2-\Delta)}{(1-\Delta)^2}}_{\det{[J^{-1}]}},
\end{equation} 
which can be determined from the trade-offs given in Refs.~\cite {yung_comparison_2024} or \cite{lu_incorporating_2021}. 

\subsection{Two-copy measurements}
For two-copy measurements, we consider the probe $\rho^{\otimes 2} = \rho\otimes\rho$, for which the SLD operators are given by $L_i^{(2)}=L_i\otimes \openone + \openone\otimes L_i$. 

For $L_\phi^{(2)}$, we have
\begin{equation}
	L_\phi^{(2)} = i(1-\Delta)\begin{pmatrix}
		0 & -e^{-i\phi} & -e^{-i\phi} & 0 \\
		e^{i\phi} & 0 & 0 & -e^{-i\phi} \\
		e^{i\phi} & 0 & 0 & -e^{-i\phi} \\
		0 & ie^{i\phi} & e^{i\phi} & 0
	\end{pmatrix} 
\end{equation}
with one choice of (unnormalised) eigenvectors and eigenvalues 
\begin{equation}
	\begin{matrix}
	0~ &0~ & -2(1-\Delta)~ & 2(1-\Delta)~ \\
	\begin{pmatrix}
		e^{-2i\phi} \\ 0 \\ 0 \\ 1
	\end{pmatrix}, & \begin{pmatrix}
		0 \\-1\\1\\0
	\end{pmatrix}, &
	\begin{pmatrix}
		-e^{-2i\phi}\\ ie^{-i\phi}\\ ie^{-i\phi}\\ 1
	\end{pmatrix},& \begin{pmatrix}
		-e^{-2i\phi}\\ -ie^{-i\phi}\\ -ie^{-i\phi}\\ 1
	\end{pmatrix}.
	\end{matrix}\label{eq:evecs}
\end{equation}
We can thus form the three projectors 
\begin{align}
	\begin{split}
		P_0 &= \frac{1}{2}\begin{pmatrix}
			1 & 0 & 0 & e^{-2i\phi} \\
			0 & 1 & -1 & 0 \\
			0 & -1 & 1 & 0 \\
			e^{2i\phi} & 0 & 0 & 1
		\end{pmatrix}, \\
		P_{-} &= \frac{1}{4}\begin{pmatrix}
			1 & ie^{-i\phi} & ie^{-i\phi} & -e^{-2i\phi} \\
			-ie^{i\phi} & 1 & 1 & ie^{-i\phi} \\
			-ie^{i\phi} & 1  & 1 & ie^{-i\phi} \\
			-e^{2i\phi} & -ie^{i\phi} & -ie^{i\phi} & 1
		\end{pmatrix}, \\
		P_{+} &= \frac{1}{4}\begin{pmatrix}
			1 & -ie^{-i\phi} & -ie^{-i\phi} & -e^{-2i\phi} \\
			ie^{i\phi} & 1 & 1 & -ie^{-i\phi} \\
			ie^{i\phi} & 1  & 1 & -ie^{-i\phi} \\
			-e^{2i\phi} & ie^{i\phi} & ie^{i\phi} & 1
		\end{pmatrix}.
	\end{split}
\end{align}

We then project $\rho^{\otimes 2}$ onto each subspace and calculate the quantum Fisher information for $\Delta$ via its SLD operator. We find a $\Delta$-dependent projected state
\begin{equation}
	P_0\rho^{\otimes2} P_0^\dagger = \frac{1}{8}\begin{pmatrix}
		2-\Delta(2-\Delta) & 0 & 0 & (2-\Delta(2-\Delta))e^{-2i\phi}  \\
		0 & \Delta(2-\Delta) & -\Delta(2-\Delta) & 0 \\
		0 & -\Delta(2-\Delta) & \Delta(2-\Delta) & 0 \\
		(2-\Delta(2-\Delta))e^{2i\phi} & 0 & 0 & 2-\Delta(2-\Delta)
	\end{pmatrix}
\end{equation}
and corresponding SLD operator
\begin{equation}
	L_\Delta^{P_0} = \begin{pmatrix}
		a_{11} & a_{12} & a_{12} & -a_{11}e^{-2i\phi} - \frac{2e^{-2i\phi}(1-\Delta)}{2-2\Delta + \Delta^2} \\
		a_{21} & a_{22} & a_{22}-\frac{2(1-\Delta)}{\Delta(2-\Delta)} & -a_{21}e^{-2i\phi} \\
		a_{21} & a_{22}-\frac{2(1-\Delta)}{\Delta(2-\Delta)} & a_{22} & -a_{21}e^{-2i\phi} \\
		-a_{11}e^{2i\phi}- \frac{2e^{2i\phi}(1-\Delta)}{2-2\Delta + \Delta^2} & -a_{12}e^{2i\phi} & -a_{12}e^{2i\phi} & a_{11}
	\end{pmatrix},
\end{equation} 
with arbitrary $a_{11}$, $a_{12}$, $a_{22}$, $a_{21}=a_{12}^\ast$. Then, the quantum Fisher information of $\Delta$ in the subspace is
\begin{equation}
	J_\Delta^{P_0} = \qtr[(P_0\rho^{\otimes2} P_0^\dagger)L_\Delta^{P_0}L_\Delta^{P_0}] = \frac{2}{\frac{1}{(1-\Delta)^2} + \Delta(2-\Delta)-1},
\end{equation}
which is non-zero. On the other hand, the equivalent quantum Fisher informations from the other subspaces are zero because the projected states $P_-\rho^{\otimes2} P_-^\dagger$ and $P_+\rho^{\otimes2} P_+^\dagger$ do not depend on $\Delta$. 

In this system, the $\Delta$-optimal measurement is not sensitive to $\phi$. To see this, we can calculate the projectors for the $\Delta$-SLD operator (at the local value $\phi=0$):
\begin{align}
	\begin{split}
		P_0^{(\Delta)} &= \frac{1}{2}\begin{pmatrix}
			1 & 0 & 0 & -1 \\
			0 & 1 & -1 & 0 \\
			0 & -1 & 1 & 0 \\
			-1 & 0 & 0 & 1
		\end{pmatrix}, \\
		P_-^{(\Delta)} &= \frac{1}{4} \begin{pmatrix}
			1 & 1 & 1 & 1 \\
			1 & 1 & 1 & 1 \\
			1 & 1 & 1 & 1 \\
			1 & 1 & 1 & 1 
		\end{pmatrix}, \\
		P_+^{(\Delta)} &=\frac{1}{4} \begin{pmatrix}
			1 & -1 & -1 & 1 \\
			-1 & 1 & 1 & -1 \\
			-1 & 1 & 1 & -1 \\
			1 & -1 & -1 & 1 
		\end{pmatrix}.
	\end{split}
\end{align}
Then, the two-copy state projected onto the eigenspaces (at $\Delta=1/2$) is
\begin{align}
	\begin{split}
		P_0^{(\Delta)}\rho^{\otimes2}P_0^{(\Delta)\dagger} &= \frac{1}{32} \begin{pmatrix}
			4-\cos(2\phi) & 0 & 0 & -4+\cos(2\phi) \\
			0 & 3 & -3 & 0 \\
			0 & -3 & 3 & 0 \\
			-4+\cos(2\phi) & 0 & 0 & 4-\cos(2\phi)
		\end{pmatrix},\\
		P_-^{(\Delta)}\rho^{\otimes2}P_-^{(\Delta)\dagger} &= \frac{1}{2}(2+\cos\phi)^2\begin{pmatrix}
			1 & 1 & 1 & 1 \\
			1 & 1 & 1 & 1 \\
			1 & 1 & 1 & 1 \\
			1 & 1 & 1 & 1 
		\end{pmatrix}, \\
		P_+^{(\Delta)}\rho^{\otimes2}P_+^{(\Delta)\dagger} &= \frac{1}{2}(2-\cos\phi)^2\begin{pmatrix}
			1 & -1 & -1 & 1 \\
			-1 & 1 & 1 & -1 \\
			-1 & 1 & 1 & -1 \\
			1 & -1 & -1 & 1 
		\end{pmatrix}.
	\end{split}
\end{align} 
Each of the projected states has zero derivative at $\phi=0$, and therefore the quantum Fisher information about $\phi$ (at $\phi=0$) for each of the projected states is zero. For a different true $\phi$ value, the projectors would be different, but the overall result would be the same.  

\subsection{Collective measurement trade-off calculation}
We calculate the multiple-copy trade-off curve numerically by calculating the Nagaoka Cram\'er--Rao bound with different weight matrices (of form $\operatorname{Diag}(a,b)$). The lower bound $\mathcal{C}(a,b)$ dictates that the variances satisfy $a V(\hat\phi) + b(\hat\Delta)\geq \mathcal{C}(a,b)$, which defines a closed half-plane in $V(\hat\phi)$--$V(\hat\Delta)$ space. By varying $a$ and $b$ (specifically their ratio), half-planes with different boundary gradients are determined. The trade-off curve is the boundary of the intersection of all such half-planes. 

To numerically determine the intersection of the half-planes, we use the \texttt{qhalf} routine \footnote{\url{https://www.qhull.org/html/qhalf.htm}} of the \texttt{Qhull} program, which implements the Quickhull algorithm for computing convex hulls \cite{barber_quickhull_1996}. The program output is the intersection points of the half-spaces, which form a lower bound to the true trade-off curve for the Nagaoka Cram\'er--Rao bound, approaching the true curve as the number of half-planes is increased. The input to \texttt{qhalf} includes values of $a$, $b$, and $\mathcal{C}(a,b)$ calculated using a semidefinite program \cite{conlon_efficient_2021}.

\section{Fock state displacement sensing details} \label{sec:fock}
We consider sensing $\theta_x$, $\theta_y$ in 
\begin{equation}
	\ket{\psi(\theta_x,\theta_y)} = D(\theta_x,\theta_y)\ket{n},
\end{equation}
where $D(\theta_x,\theta_y) = \exp(z\hat{a}^\dagger -z^*\hat{a}) = \exp(i\theta_x\hat{y} + i\theta_y \hat{x})$, ($z= (\theta_x+i\theta_y)/\sqrt{2}$). The derivatives are
\begin{equation}
	\frac{\partial}{\partial\theta_x}\ket{\psi(\theta_x,\theta_y)} = \frac{1}{\sqrt{2}}\left(\hat{a}^\dagger - \hat{a}\right) \ket{\psi(\theta_x,\theta_y)}, 
\end{equation}
\begin{equation}
	\frac{\partial}{\partial\theta_y}\ket{\psi(\theta_x,\theta_y)} = \frac{i}{\sqrt{2}}\left(\hat{a}^\dagger + \hat{a}\right) \ket{\psi(\theta_x,\theta_y)}.
\end{equation}
If we set the true values $\theta_x=\theta_y=0$, then $\ket{\psi} = \ket{n}$ and 
\begin{equation}
	\ket{\partial_x\psi} = \frac{1}{\sqrt{2}}(\hat{a}^\dagger-\hat{a})\ket{n} = \frac{1}{\sqrt{2}}\left(\sqrt{n+1}\ket{n+1}-\sqrt{n}\ket{n-1}\right),
\end{equation}
\begin{equation}
	\ket{\partial_y\psi} = \frac{i}{\sqrt{2}}(\hat{a}^\dagger+\hat{a})\ket{n} = \frac{i}{\sqrt{2}}\left(\sqrt{n+1}\ket{n+1}+\sqrt{n}\ket{n-1}\right).
\end{equation}
The system is then effectively three-dimensional, and for $n>0$ we can equivalently write the matrices
\begin{equation}
	\rho = \begin{pmatrix}
		0 & 0 & 0 \\
		0 & 1 & 0 \\
		0 & 0 & 0
	\end{pmatrix}, \quad \partial_x\rho = \frac{1}{\sqrt{2}}\begin{pmatrix}
		0 & -\sqrt{n} & 0 \\
		-\sqrt{n} & 0 & \sqrt{n+1} \\
		0 & \sqrt{n+1} & 0
	\end{pmatrix}, \quad \partial_y\rho = \frac{i}{\sqrt{2}}\begin{pmatrix}
		0 & \sqrt{n} & 0 \\
		-\sqrt{n} & 0 & -\sqrt{n+1} \\
		0 & \sqrt{n+1} & 0
	\end{pmatrix}.
\end{equation}
For a pure state problem, the quantum Fisher information is $J = 4\re{\left(\braket{\partial_i\psi}{\partial_j\psi}-\braket{\partial_i\psi}{\psi}\braket{\psi}{\partial_j\psi}\right)}$. We find
\begin{equation}
	J = 4\left(n+\frac{1}{2}\right)\openone_2. 
\end{equation} 
For the trade-off, for this pure state problem with diagonal quantum Fisher information, we can use the trade-off from Ref.~\cite{lu_incorporating_2021} with the incompatibility coefficient $c=1/(2n+1)$. 

\subsection{Prioritised measurement} 
Since we are working with a rank-deficient problem, the SLD operators are not unique. In particular, we can have
\begin{equation}
	L_x = \begin{pmatrix}
		a & -\sqrt{2}\sqrt{n} & b+ic \\
		-\sqrt{2}\sqrt{n} & 0 & \sqrt{2}\sqrt{n+1} \\
		b-ic & \sqrt{2}\sqrt{n+1} & d 
	\end{pmatrix} \label{eq:Lx}
\end{equation}
for arbitrary real $a$, $b$, $c$, $d$. It is, however, still possible to compute the eigenvectors of $L_x$ and calculate the classical Fisher information with respect to $\theta_y$ of the corresponding projective measurement. This can then be maximised with respect to the real numbers. We find a (non-unique) solution $a=b=d=0$ and $c=1$, with Fisher information $F_{yy}=16/3$ (for $n=1$), which saturates the trade-off curve. 

\section{Quantinuum Experiment Details}
For the measurement saturating the quantum Cram\'er--Rao bound for $\phi$, we use the POVM with elements:
\begin{equation}
	\Pi_i = \ketbra{\psi_i}, \quad \ket{\psi_1} = \frac{1}{2}\begin{pmatrix}
		1 \\
		-i\\
		-i\\
		-1
	\end{pmatrix},\ket{\psi_2} = \frac{1}{2}\begin{pmatrix}
		1 \\
		i\\
		i\\
		-1
	\end{pmatrix}, \ket{\psi_3} = \frac{1}{\sqrt{2}}\begin{pmatrix}
		0 \\
		-1\\
		1\\
		0
	\end{pmatrix},
	\ket{\psi_4} = \frac{1}{\sqrt{2}}\begin{pmatrix}
		1 \\
		0\\
		0\\
		1
	\end{pmatrix}, \label{eq:phipovm}
\end{equation}
which is locally optimal for the true value $\phi=0$ (is is the projection measurement in an eigenbasis of $L_\phi^{(2)}$). The parameters are estimated using the coefficients (for local value $\Delta=1/2$)
\begin{align}
	\hat{\phi}_1 &= -2, \ \hat{\phi}_2 = 2, \ \hat{\phi}_3 = 0, \ \hat{\phi}_4 = 0, \\
	\hat{\Delta}_1 &= 0, \ \hat{\Delta}_2 = 0, \hat{\Delta}_3 = 5/2, \hat{\Delta}_4 = -3/2,
\end{align}
such that 
\begin{equation}
	\hat{\phi} = \sum_{i=1}^4 \hat{\phi}_i p_i, \quad \hat{\Delta} = \sum_{i=1}^4 \hat{\Delta}_i p_i + 1/2,
\end{equation}
where $p_i$ is the outcome frequency of POVM element $\ketbra{\psi_i}$. 

\begin{figure}
	\centering
	\includegraphics[width=\linewidth]{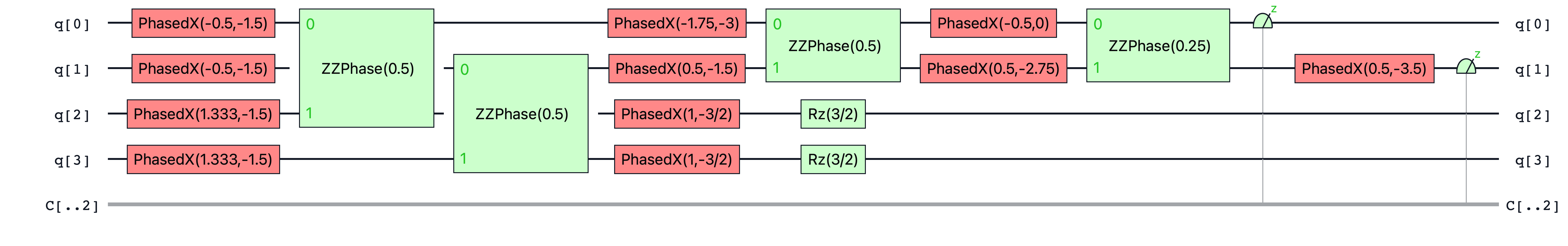}
	\caption{Circuit diagram for the computation implemented on the Quantinuum H1-1 device. The circuit prepares two copies of a dephased qubit (using ancillas) and then implements an entangling measurement. Diagram generated using \texttt{pytket} (\url{https://docs.quantinuum.com/tket/api-docs/index.html}).}
	\label{fig:circ}
\end{figure}

\begin{figure}
	\centering
	\includegraphics[width=0.8\linewidth]{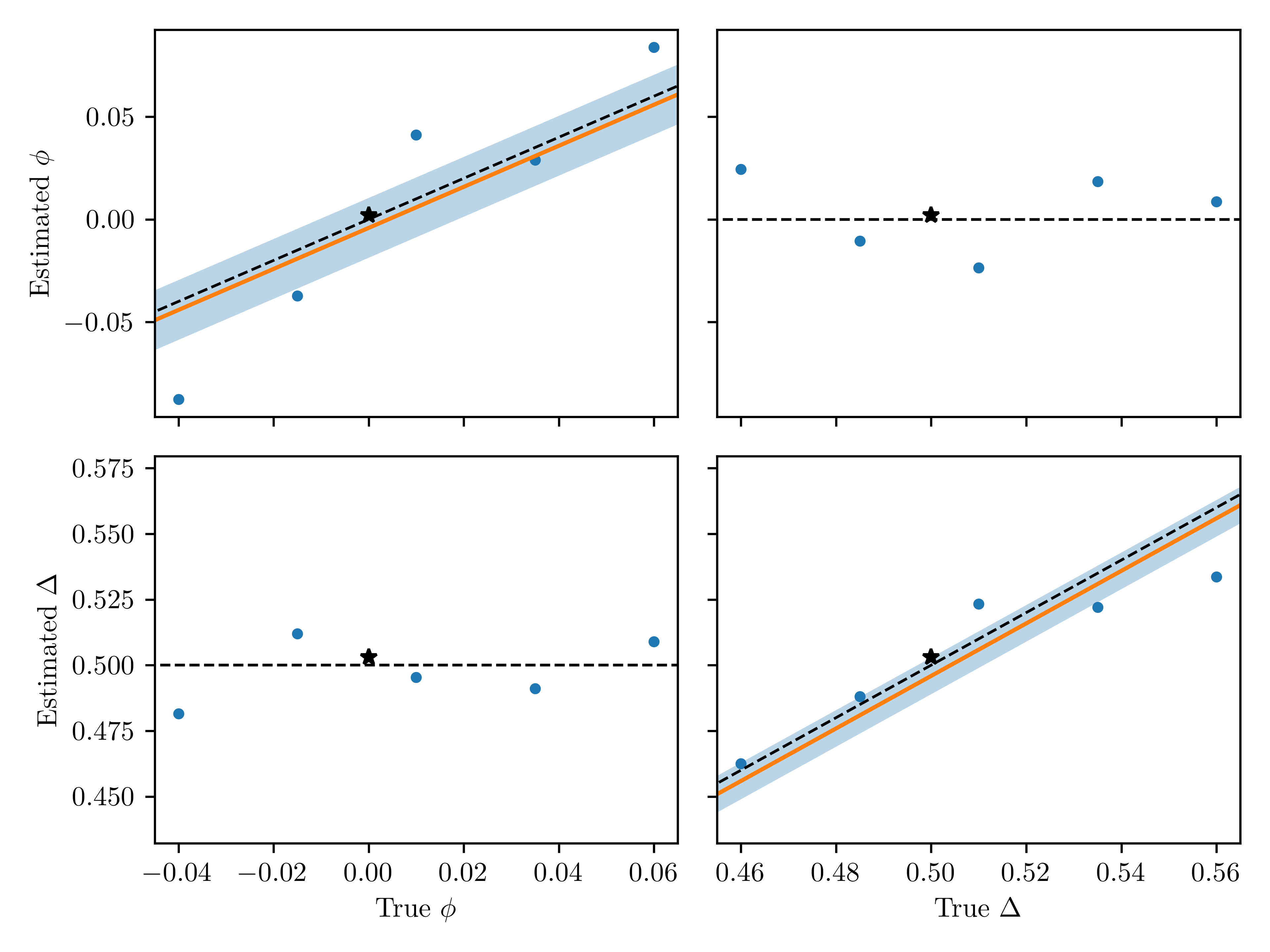}
	\caption{Estimated parameters with different true values from Quantinuum experiment. The stars denotes the results from the main run with $\phi=0$ and $\Delta=1/2$ (for which the measurement is optimised). The black dashed lines denote ``estimated = true''. The orange lines denote a best-fit of ``estimated = true + offset'', and the blue shaded regions denote one standard deviation of the fitted offset.}
	\label{fig:calib}
\end{figure}

The quantum circuit is presented in Fig.~\ref{fig:circ} and was implemented on the Quantinuum H1-1 trapped-ion device. After the state preparation, the circuit implements a unitary that is the inverse of the unitary with columns as the vectors in Eq.~\eqref{eq:phipovm}, so that measurement in the computational basis implements the POVM. In total, \SI{10000}{} shots were collected at each of the true values $\phi\in\{-0.04,-0.015,0.01,0.035,0.06\}$ and $\Delta=0.5$, $\phi=0$ and $\Delta\in\{0.46,0.485,0.51,0.535,0.56\}$, and the main run with $\phi=0$ and $\Delta=0.5$. The results are presented in Fig.~\ref{fig:calib}, demonstrating the unbiasedness of the measurement.  

The mean squared errors for the parameter estimates in the main run were calculated by bootstrapping the experimental data. Here, the experimental data were resampled to simulate \SI{10000}{} runs, each with 200 shots to estimate the variation in the estimates. This was then repeated 500 times to obtain the average and uncertainty of the mean squared error (reported in the Figure in the main text).   

For the measurements saturating the two-copy Nagaoka--Hayashi Cram\'er--Rao bound with different weights, we use numerically optimised POVMs. The weight matrices are of the form $W = \operatorname{Diag}(w,2-w)$ with $w\in\{1.8,1.4,1,0.6,0.2\}$. From left to right along the curve in Fig.~1 in the main text (excluding the left-most, which is the same as above and corresponds to the limit $w\rightarrow 2$), the projectors are
\begin{align}
	\ket{\psi_1} &= \begin{pmatrix}
		0.5033 \\
		-0.0569+0.4934i\\
		-0.0569+0.4934i \\
		-0.4901-0.1145i 
	\end{pmatrix}, \ket{\psi_2} = \begin{pmatrix}
		0 \\
		1/\sqrt{2} \\
		-1/\sqrt{2} \\
		0
	\end{pmatrix}, \ket{\psi_3} = \begin{pmatrix}
		0.7024 \\
		0.0815 \\
		0.0815 \\
		0.7024
	\end{pmatrix}, \ket{\psi_4} = \begin{pmatrix}
		0.5033 \\
		-0.0569-0.4934i \\
		-0.0569-0.4934i \\
		-0.4901+0.1145i
	\end{pmatrix} \\
	\ket{\psi_1} &= \begin{pmatrix}
		 0.5135 \\
		 -0.1109 + 0.4733i \\
		 -0.1109 + 0.4733i \\
		 -0.4601 - 0.2282i
	\end{pmatrix}, \ket{\psi_2} = \begin{pmatrix}
		0 \\
		1/\sqrt{2} \\
		-1/\sqrt{2} \\
		0
	\end{pmatrix}, \ket{\psi_3} = \begin{pmatrix}
		0.6874 \\
		0.1657\\
		0.1657 \\
		0.6874 
	\end{pmatrix}, \ket{\psi_4} = \begin{pmatrix}
		0.5135\\
		-0.1109 - 0.4733i\\
		-0.1109 - 0.4733i \\
		-0.4601 + 0.2282i
	\end{pmatrix} \\
	\ket{\psi_1} &= \begin{pmatrix}
		 0.5245  \\
		 -0.1434 + 0.4520i \\
		 -0.1434 + 0.4520i \\
		 -0.4287 - 0.3023i
	\end{pmatrix}, \ket{\psi_2} = \begin{pmatrix}
		0 \\
		1/\sqrt{2} \\
		-1/\sqrt{2} \\
		0
	\end{pmatrix}, \ket{\psi_3} = \begin{pmatrix}
		0.6706 \\
		0.2243 \\
		0.2243 \\
		0.6706
	\end{pmatrix}, \ket{\psi_4} = \begin{pmatrix}
		0.5245 \\
		-0.1434 - 0.4520i \\
		-0.1434 - 0.4520i \\
		-0.4287 + 0.3023i
	\end{pmatrix} \\
	\ket{\psi_1} &= \begin{pmatrix}
		 0.5382 \\
		 -0.1697 + 0.4260i \\
		 -0.1697 + 0.4260i \\
		 -0.3907 - 0.3701i
	\end{pmatrix}, \ket{\psi_2} = \begin{pmatrix}
		0 \\
		1/\sqrt{2} \\
		-1/\sqrt{2} \\
		0
	\end{pmatrix}, \ket{\psi_3} = \begin{pmatrix}
		0.6486 \\
		0.2817\\
		0.2817\\
		0.6486 
	\end{pmatrix}, \ket{\psi_4} = \begin{pmatrix}
		0.5382 \\
		-0.1697 - 0.4260i\\
		-0.1697 - 0.4260i\\
		-0.3907 + 0.3701i
	\end{pmatrix} \\
	\ket{\psi_1} &= \begin{pmatrix}
		 0.5620 \\
		 -0.1960 + 0.3817i\\
		 -0.1960 + 0.3817i \\
		 -0.3276 - 0.4567i
	\end{pmatrix}, \ket{\psi_2} = \begin{pmatrix}
		0 \\
		1/\sqrt{2} \\
		-1/\sqrt{2} \\
		0
	\end{pmatrix}, \ket{\psi_3} = \begin{pmatrix}
		0.6068 \\
		0.3630 \\
		0.3630 \\
		0.6068
	\end{pmatrix}, \ket{\psi_4} = \begin{pmatrix}
		0.5620 \\
		-0.1960 - 0.3817i \\
		-0.1960 - 0.3817i \\
		-0.3276 + 0.4567i
	\end{pmatrix}		
\end{align}
The corresponding estimator coefficients are
\begin{align}
	\hat{\theta} &= \begin{pmatrix}
		2.1375 & 0 & 0 & -2.1375 \\
		0.3581 & 1.8459 & -1.3648 & 0.3581 
	\end{pmatrix} \quad [w=1.8]\\
	\hat{\theta} &= \begin{pmatrix}
		2.3392 & 0 & 0 & -2.3392 \\
		0.5882 & 1.3291 & -1.1170 & 0.5882 
	\end{pmatrix} \quad [w=1.4]\\
	\hat{\theta} &= \begin{pmatrix}
		2.5323 & 0 & 0 & -2.5323 \\
		0.6897 & 1.0688 & -0.9538 & 0.6897 
	\end{pmatrix} \quad [w=1]\\
	\hat{\theta} &= \begin{pmatrix}
		2.7856 & 0 & 0 & -2.7856 \\
		0.7582 & 0.8777 & -0.8181 & 0.7582 
	\end{pmatrix} \quad [w=0.6]\\
	\hat{\theta} &= \begin{pmatrix}
		3.3250 & 0 & 0 & -3.3250 \\
		0.8193 & 0.6937 & -0.6743 & 0.8193
	\end{pmatrix}\quad [w=0.2]
\end{align}
where the first row of each matrix is the coefficients for $\hat{\phi}$, and the second row is the coefficients for $\hat{\Delta}$. 

\section{Proof of Theorem 1}

In the main text, we present the following theorem

\begin{theorem}
	Let $\rho(\theta_p,\theta_o)$ be a regular full-rank two-parameter quantum statistical model with linearly independent parameters. Let $O_p$ be the optimal observable for estimating $\theta_p$. Then, $\theta_o$ can be estimated whilst optimally estimating $\theta_p$ if and only if the projectors $\{P_j\}$ of $O_p$ can be chosen such that $\partial_{\theta_o}\qtr[\rho_\theta P_j]\neq 0$ for some $j$. Furthermore, this is equivalent to the classical Fisher information of the projective measurement $\{P_j\}$ being positive-definite. 
\end{theorem}

For clarity, we assume the following regularity conditions: $\btheta \rightarrow \rho(\btheta)$ is a one-to-one map, from an open set of $\mathbb{R}^2$ to the Hilbert space, that is smooth, such that sufficiently many derivatives exist. We further assume that the derivatives, $\partial\rho/\partial \theta_i$, are linearly independent. These assumptions ensure that the Cram\'er--Rao bounds hold. 

To prove the theorem, we will assume that the parameters are locally orthogonal and the quantum Fisher information is diagonal. We do this without loss of generality, since given a model $\rho(\theta_1,\theta_2)$, it is always possible to find a new parameterisation $\tilde\theta_1,\tilde\theta_2$ such that $\tilde\theta_1=\theta_1$ (i.e., the prioritised parameter is unchanged) and the quantum Fisher information is diagonal \cite{suzuki_quantum_2020}:
\begin{equation}
	J(\rho_{\tilde\theta}) = \operatorname{diag}(J_{11}(\rho_{\tilde{\theta}}),J_{22}(\rho_{\tilde\theta}))
\end{equation} 
\begin{equation}
	J_{11}(\rho_{\tilde\theta}) = J_{11}(\rho_\theta)-J_{12}(\rho_\theta)(J_{22}(\rho_\theta))^{-1}J_{21}(\rho_\theta).
\end{equation}
Similarly, the classical Fisher information can always be made diagonal by a suitable reparameterisation. Since the reparameterisation is simply a mathematical description that does not change the underlying physical model, if prioritised parameter estimation (about $\theta_1$) is possible in the orthogonal parameterisation, it is possible in the original parameterisation. Therefore, for clarity, we assume that the Fisher information matrices are diagonal. 

Furthermore, the following property is useful:

($\star$): \textit{Score-function-based estimator.} Given a statistical model $\{p_\theta(x)~|~\theta \in \Theta\}$, let $s_i(x,\theta) = \partial_{\theta_i}\log p_\theta(x)$ be a score function. Then, the estimator
\begin{equation}
	\hat\theta_i^*(x) := \theta_i + \sum_{j=1}^n[F(p_\theta)^{-1}]_{ji}s_j(x,\theta)
\end{equation}
is locally unbiased at $\theta$ and its mean squared error matrix is $V_\theta[\hat\theta^*] = F(p_\theta)^{-1}$ \cite{suzuki_quantum_2020}. That is, the estimator saturates the classical Cram\'er--Rao bound.

\textit{Proof of Theorem 1:} We can reframe the theorem as: the following conditions are equivalent:
\begin{itemize}
	\item (i) Prioritised estimation of $\theta_p$ is possible;
	\item (ii) There exists $\{P_j\}$ such that $\partial_{\theta_o}\qtr[\rho_\theta P_j]\neq 0$ for some $j$;
	\item (iii) There exists $\{P_j\}$ such that $F(\{P_j\})\succ 0$. 
\end{itemize}

Note that (ii) implies that $\qtr[\rho_\theta P_j]\neq 0$ also. 

The implication [(i) $\Longrightarrow$ (ii)] is clear, because $\theta_o$ cannot be estimated if the probability distribution $\{p_\theta(j) =\qtr[\rho_\theta P_j]\}$ does not depend on $\theta_o$. 

The equivalence [(ii) $\Longleftrightarrow$ (iii)] is a computation following the definition of the classical Fisher information. (The assumption of diagonal Fisher matrices means that $F_{22}>0$ (the element for $\theta_o$) implies $F\succ 0$.) 

Finally, [(i) $\Longleftarrow$ (ii)], can be shown using the score-function-based estimator ($\star$). If (ii) is satisfied, we can explicitly construct a locally unbiased estimator for $\theta_o$, and the (11) component of the mean squared error is equal to $[J(\rho_\theta)^{-1}]_{11}$, the requirement for prioritised estimation.   

When the projectors $\{P_j\}$ are not unique (because the SLD operator has degenerate eigenvalues), we need to check the different choices for the projectors. 

For rank-deficient models, the theorem does not always hold, as it is possible that different optimal observables $O_p$ provide different information about $\theta_o$. In particular, the model in Sec.~\ref{sec:fock} is an example of this, where prioritised estimation is possible, but a choice of $O_p=L_x$ from Eq.~\eqref{eq:Lx} with $a=b=c=d=0$ yields probabilities that do not depend on $\theta_y$.

\bibliography{bib.bib}